\documentclass[aps,11pt,nofootinbib,prd,longbibliography]{revtex4-1}

\usepackage{feynmp-auto}
\usepackage{amsmath}
\usepackage[utf8x]{inputenc} 

\usepackage{bbm}
\usepackage{soul}	
\usepackage{xcolor}

\usepackage{epsfig}
\usepackage{color}
\usepackage{slashed}
\usepackage{comment}
\usepackage{epstopdf}

\usepackage{array}
\usepackage{booktabs}
\usepackage{mathrsfs}
\usepackage[euler]{textgreek}
\usepackage{amsmath}
\usepackage{amsthm}
\usepackage{amsfonts}
\usepackage{amssymb}
\usepackage{graphicx}
\graphicspath{ {Figures/} }
\usepackage[font={small}]{caption}
\usepackage{float}
\usepackage{multirow}
\usepackage[export]{adjustbox}
\usepackage[hang, flushmargin,bottom]{footmisc} 
\usepackage{hyperref}
\hypersetup{
     colorlinks   = true,
     citecolor    = blue
}

\usepackage{placeins}
\usepackage{enumitem}

\usepackage{tikz-feynman} 
\tikzfeynmanset{compat=1.0.0}

\usepackage{slashed}
\usepackage{bbm}
\usepackage{appendix}

\usepackage{titlesec}
\titleformat{\chapter}[display]
  {\normalfont\LARGE\bfseries}
  {\chaptertitlename\ \thechapter}{5pt}{\LARGE}
  \titlespacing*{\chapter}{0pt}{-20pt}{35pt}
\usepackage{bigstrut}
\usepackage{pst-node}
\usepackage{pstricks}

\usepackage{physics}
\usepackage{mathtools}
\usepackage{setspace}
\usepackage{fancyhdr}
\usepackage[makeroom]{cancel}

\newcommand{\be}{\begin{equation}}
\newcommand{\ee}{\end{equation}}
\newcommand{\bes}{\begin{equation*}}
\newcommand{\ees}{\end{equation*}}

\usepackage{hyperref}
\hypersetup{%
  colorlinks = true,
  linkcolor  = black
}
\usepackage{soul}

\newcommand{\beq}{\begin{equation}}
\newcommand{\eeq}{\end{equation}}

\newcommand{\SU}{\,{\rm SU}}
\newcommand{\U}{\,{\rm U}}

\usepackage[normalem]{ulem}

\DeclareUnicodeCharacter{2212}{-}

\AtBeginDocument{%
    \newwrite\bibnotes
    \def\bibnotesext{Notes.bib}
    \immediate\openout\bibnotes=\jobname\bibnotesext
    \immediate\write\bibnotes{@CONTROL{REVTEX41Control}}
    \immediate\write\bibnotes{@CONTROL{%
    apsrev41Control,author="08",editor="1",pages="1",title="0",year="1"}}
     \if@filesw
     \immediate\write\@auxout{\string\citation{apsrev41Control}}%
    \fi
}%

\usepackage{amsfonts}
\usepackage{graphicx}
\usepackage{tcolorbox}
\usepackage{marvosym} 
\usepackage{tikz}
\usepackage{tabulary}
\usepackage{pgfplots}
\usepackage{subfig}
\usepackage{amsmath}
\usepackage{slashed}
\usepackage{mathtools}
\usepackage{nccmath}
\usepackage{verbatim}
\usepackage{color}
\usepackage{amsmath, amsthm, amssymb, amsfonts}
\usepackage{array}
\newcolumntype{P}[1]{>{\centering\arraybackslash}p{#1}}
\usepackage{enumerate}
\usepackage{physics}
\usepackage[font={small, up}]{caption}
\graphicspath{{./images/}}
\usepackage{appendix}

\usepackage{hyperref}	

\begin{document}
\title{\Large {\bf{Two-Higgs-Doublet Model and Quark-Lepton Unification}}}
\author{Pavel Fileviez P\'erez$^{1}$, Elliot Golias$^{1}$, Alexis D. Plascencia$^{2}$}
\affiliation{$^{1}$Physics Department and Center for Education and Research in Cosmology and Astrophysics (CERCA), Case Western Reserve University, Cleveland, OH 44106, USA \\
$^{2}$INFN, Laboratori Nazionali di Frascati, C.P. 13, 100044 Frascati, Italy}
\email{pxf112@case.edu, elliot.golias@case.edu, alexis.plascencia@lnf.infn.it}
\vspace{1.5cm}

\begin{abstract}
We study the Two-Higgs-Doublet Model predicted in the minimal theory for quark-lepton unification that can describe physics at the low scale. We discuss the relations among the different decay widths of the new Higgs bosons and study their phenomenology at the Large Hadron Collider. As a result of matter unification, this theory predicts a correlation between the decay widths of the heavy Higgs bosons into tau leptons and bottom quarks. We point out how to probe this theory using these relations and discuss the relevant flavor constraints.
\end{abstract}

\maketitle
\hypersetup{linkcolor=blue}
\section{INTRODUCTION}
%
After the discovery of the Standard Model (SM) Brout-Englert-Higgs boson at the Large Hadron Collider (LHC) the question remains open of whether there exist more particles within reach of experiments. 
One simple possibility is to have a second Higgs doublet carrying the same quantum numbers as the SM Higgs. The Two-Higgs-Doublet Model (2HDM) is predicted in different extensions beyond the SM and can provide a framework for dark matter, spontaneous CP violation and baryogenesis. However, there is a large freedom in the parameter space of this simple model and it is difficult to make unique predictions that can be tested by different experiments. For reviews on the 2HDM we refer the reader to Refs.~\cite{Gunion:1989we,Branco:2011iw}.

The idea of matter unification proposed by J. Pati and A. Salam~\cite{Pati:1974yy} remains one of the best ideas for theories for physics beyond the Standard Model. In this context, the SM quarks and leptons live in the same 
representations and the theory predicts the existence of right-handed neutrinos needed for the seesaw mechanism for neutrino masses.  The Pati-Salam symmetry, $\SU(4)_C$, must be broken around the canonical seesaw 
scale, $M_R \sim 10^{14}$ GeV, if neutrino masses are generated through the Type-I seesaw mechanism because the theory predicts similar values for the Dirac neutrino mass matrix and the mass matrix for the up-quarks.
Flavor violating processes also impose a non-trivial bound on the scale of new physics.
The vector leptoquark, $X_\mu \sim (\mathbf{3}, \mathbf{1}, 2/3)$, predicted by the $\SU(4)_C$ symmetry must be generically heavy, $M_X \gtrsim 10^3$ TeV, in order to satisfy the experimental bounds 
on the lepton flavor number violating rare Kaon decays, i.e. $K_L \to e^\pm \mu^\mp$. It is important to mention that this bound can be relaxed since we do not know the values of the mixing between quarks and leptons entering in these predictions.

A simple theory for quark-lepton unification at the low scale was proposed in Ref.~\cite{FileviezPerez:2013zmv}. This theory is based on $\SU(4)_C \otimes \SU(2)_L \otimes \U(1)_R$, which is the minimal gauge group that can be used to unify matter, and neutrino masses are generated through the inverse seesaw mechanism~\cite{Mohapatra:1986aw,Mohapatra:1986bd} so that the breaking of $\SU(4)_C$ can occur at the low scale. The minimal way to break the degeneracy between the masses for the down-type quarks and the charged leptons is to introduce a scalar $\Phi \sim (\mathbf{15},\mathbf{2},1/2)$ which contains a Higgs doublet with the same quantum numbers as the doublet in the SM, $H_2 \sim (\mathbf{1},\mathbf{2},1/2)$. Therefore, this theory predicts a simple Higgs sector with two Higgs doublets but the $\SU(4)_C$ symmetry predicts unique relations between the Yukawa interactions for these Higgs bosons. 

In Ref.~\cite{Perez:2021mgz} we recently pointed out relations between the decay widths of the scalar leptoquarks and the new Higgs bosons that can be used to probe the idea of quark-lepton unification. In this article, we study the phenomenology and study how the relations between the Higgs decay widths are realized in scenarios that are compatible with LHC searches and experimental constraints on flavor violating observables. We demonstrate that the relation between the heavy CP-even and CP-odd Higgs decays can be used to test the idea of quark-lepton unification. In the case when there is non-flavor violation one finds the simple relations:
$\Gamma(H \to \bar{\tau} \tau) = 3 \, \Gamma(H \to \bar{b} b)$ and $\Gamma(A \to \bar{\tau} \tau) = 3 \, \Gamma(A \to \bar{b} b)$  for small values of $\tan \beta$, and
$\Gamma(H \to \bar{\tau} \tau) = \Gamma(H \to \bar{b} b) /3$  and $\Gamma(A \to \bar{\tau} \tau) =  \Gamma(A \to \bar{b} b)/3$ for large values of $\tan \beta.$
We also discuss the properties of the charged Higgs decays and the LHC constraints to understand the testability of the theory using the Higgs decays.
The constraints coming from $K-\bar{K}$ mixing and the lepton number violating process $\mu \to e \gamma$ are discussed in detail.

This article is structured as follows: in Section~\ref{sec:theory}, we overview the minimal theory of quark-lepton that can live at the low scale. In Section~\ref{sec:higgssector}, we discuss the Higgs sector of the theory. In Section~\ref{sec:decays}, we study the different decay channels of the new Higgs bosons and find relations among them predicted from quark-lepton unification. In Section~\ref{sec:LHC}, we study the production of the new scalars at the LHC study the constraints from current searches by the CMS and ATLAS collaborations. In Section~\ref{sec:flavor}, we study the constrains from flavor-violating observables and propose an ansatz for the Yukawa couplings motivated by quark-lepton unification. We summarize our results in Section~\ref{sec:summary}.
%
\section{MINIMAL THEORY FOR QUARK-LEPTON UNIFICATION}
\label{sec:theory}
The minimal theory for quark-lepton unification that can describe physics at the TeV scale was proposed in Ref.~\cite{Perez:2013osa}.
This theory is based on the gauge symmetry, $${\cal G}_{QL}=\SU(4)_C \otimes \SU(2)_L \otimes \U(1)_R,$$ and the SM matter fields are unified in the following representations:
\begin{eqnarray}
F_{QL} &=&
\left(
\begin{array}{cccc}
u_r & u_g & u_b  & \nu 
\\
d_r & d_g & d_b  & e
\end{array}
\right) \sim (\mathbf{4}, \mathbf{2}, 0), \\[1ex]
F_u &=&
\left(
\begin{array}{cccc}
u^c _r & u^c_g & u^c_b & \nu^c
\end{array}
\right) \sim (\mathbf{\bar{4}}, \mathbf{1}, -1/2), 
\\[1ex]
 F_d &=&
\left(
\begin{array}{cccc}
d^c_r & d^c_g & d^c_b & e^c
\end{array}
\right) \sim (\mathbf{\bar{4}}, \mathbf{1}, 1/2).
\end{eqnarray}
In this context the leptons can be understood as the fourth color of the fermions.

The Lagrangian of this theory can be written as
\begin{eqnarray}
\mathcal{L} _{421} &=& - \frac{1}{2}  {\rm Tr} (F_{\mu \nu} F^{\mu \nu}) - \frac{1}{2}  {\rm Tr} (W_{\mu \nu} W^{\mu \nu})  - \frac{1}{4}  B_{\mu \nu} B^{\mu \nu}  \nonumber \\
& & +  i \bar{F}_{QL} \slashed{D} F_{QL} +  i \bar{F}_{u} \slashed{D} F_{u} +  i \bar{F}_{d} \slashed{D} F_{d}  +  \mathcal{L}_Y - V(H,\chi,\Phi),
\end{eqnarray}
where $F_{\mu \nu}= \partial_\mu A_\nu -  \partial_\nu A_\mu + i g_4 [A_\mu, A_\nu]$ is the strength tensor for the $\SU(4)_C$ gauge fields, $A_\mu \sim (\mathbf{15},\mathbf{1},0)$.
$W_{\mu \nu}= \partial_\mu W_\nu -  \partial_\nu W_\mu + i g_2 [W_\mu, W_\nu]$ is the strength tensor for the $\SU(2)_L$ gauge fields, $W_\mu \sim (\mathbf{1},\mathbf{3},0)$, 
and for the $\U(1)_R$ gauge field, $B_\mu \sim (\mathbf{1},\mathbf{1},0)$, we have $B_{\mu \nu}= \partial_\mu B_\nu -  \partial_\nu B_\mu$.  
See Ref.~\cite{Perez:2021mgz} for the full expression of the scalar potential $V(H,\chi,\Phi)$.
 The covariant derivatives for the 
fermionic fields are given by
\begin{eqnarray}
\slashed{D} F_{QL} &=& \gamma^\mu (\partial_\mu + i g_4 A_\mu + i g_2 W_\mu) F_{QL}, \\
\slashed{D} F_{u} &=&  \gamma^\mu (\partial_\mu - i g_4 A_\mu^T - \frac{i}{2}  g_1 B_\mu) F_{u},   \\
\slashed{D} F_{d} &=&  \gamma^\mu (\partial_\mu - i g_4 A_\mu^T + \frac{i}{2}  g_1 B_\mu) F_{d}.
\end{eqnarray}
The Yukawa interactions for the charged fermions can be written as
\begin{align}
	-\mathcal{L}_Y \supset Y_1 F_{QL} F_u H_1 + Y_2 F_{QL}  F_u \Phi + Y_3 H_1^\dagger F_{QL} F_d + Y_4 \Phi^\dagger F_{QL} F_d +  {\rm h.c.},
\end{align}
where $H_1 \sim (\mathbf{1},\mathbf{2},1/2)$ and $\Phi \sim (\mathbf{15},\mathbf{2},1/2)$ are needed to generate fermion masses in a consistent manner. 
The $\Phi$ field contains a second Higgs doublet $H_2$ that is coupled to all the SM fermions
\beq
\Phi = 
\left(
\begin{array} {cc}
\Phi_8 & \Phi_3  \\
\Phi_4 & 0  \\
\end{array}
\right) + \sqrt{2} \, T_4 \ H_2 \sim (\mathbf{15}, \mathbf{2}, 1/2),
\eeq
where $T_4$ is one of the generators of $\SU(4)_C$ and it is normalized as
$
T_4 =
\frac{1}{2 \sqrt{6}} \rm{diag} (1,1,1,-3).
$

The small neutrino masses can be generated while allowing the $\SU(4)_C$ symmetry to be broken at the low scale using the inverse seesaw mechanism.
\begin{align}
	-\mathcal{L} \supset Y_5 F_u \chi S + \frac{1}{2} \mu S S + {\rm h.c.} \, ,
\end{align}
with three copies of SM fermionic singlets $S \sim (\mathbf{1}, \mathbf{1}, 0)$. The mass matrix for neutrinos in the basis $(\nu, \nu^c, S)$ is 
\begin{align}
	\begin{pmatrix} \nu & \nu^c & S\end{pmatrix}
	\begin{pmatrix} 0 & M_\nu^D & 0 \\ (M_\nu^D)^T & 0 & M_\chi^D \\ 0 & (M_\chi^D)^T & \mu \end{pmatrix}
	\begin{pmatrix} \nu \\ \nu^c \\ S\end{pmatrix},
\end{align}
where $M_\chi^D = Y_5 v_\chi / \sqrt{2}$. The light neutrino masses are given by
\begin{align}
m_\nu \approx \mu \left( \frac{M_\nu^D}{M_\chi^D}\right)^2,
\end{align}
where $M_\chi^D \gg M_\nu^D \gg \mu$, and the heavy neutrinos form a pseudo-Dirac pair.
Notice that the symmetry $\SU(4)_C \otimes \U(1)_R$ is broken to $\SU(3)_C \otimes \U(1)_Y$ once the Higgs, $\chi \sim (\mathbf{{4}}, \mathbf{1}, 1/2)$, 
acquires the vacuum expectation value $v_\chi$. Since $v_\chi \gg v$ we can safely neglect the mixing between $\chi$ and the Higgs bosons in the $\SU(2)_L$ doublets.
For more details about this simple theory for quark-lepton unification at the low-scale see Refs.~\cite{Faber:2018qon,Faber:2018afz,Perez:2021ddi,Perez:2021mgz,Perez:2022ouu}.
%
\section{HIGGS SECTOR}
\label{sec:higgssector}
The scalar sector in the theory is a special case of the general 2HDM in which both Higgs doublets are coupled to quarks and leptons, this is commonly referred in the literature as the type-III 2HDM. Nevertheless, since the theory arises from quark-lepton unification there are only four independent Yukawa couplings defining the interactions between the Higgs doublets and the Standard Model fermions:
\begin{align}
	-\mathcal{L} &= \bar{u}_R \left( Y_1^T H_1 + \frac{1}{2\sqrt{3}} Y_2^T H_2\right)Q_L + \bar{N}_R \left( Y_1^T H_1 - \frac{\sqrt{3}}{2} Y_2^T H_2 \right) \ell_L \nonumber \\
	&+ \bar{d}_R \left( Y_3^T H_1^{\dagger} + \frac{1}{2\sqrt{3}} Y_4^T H_2^{\dagger} \right) Q_L + \bar{e}_R \left( Y_3^T H_1^{\dagger} -\frac{\sqrt{3}}{2} Y_4^T H_2^{\dagger}\right) \ell_L + {\rm h.c.} \, ,
\end{align}
in the first line above, the contraction of $\SU(2)_L$ indices is implicit, e.g. $H_1 Q_L = \varepsilon^{ab} H_1^b Q_L^a $. The doublets are given by $H_1^T=(H_1^+,(v_1 + H_1^0 + i A_1^0)/\sqrt{2})$ and correspondingly for $H_2$. After electroweak symmetry breaking the fermion mass matrices read as
\begin{align}
	M_U &= Y_1\frac{v_1}{\sqrt{2}} + \frac{1}{2\sqrt{3}} Y_2 \frac{v_2}{\sqrt{2}}, 	\hspace{0.8cm} M_D = Y_3 \frac{v_1}{\sqrt{2}} + \frac{1}{2\sqrt{3}} Y_4 \frac{v_2}{\sqrt{2}}, \nonumber \\
	M_{\nu}^D &= Y_1 \frac{v_1}{\sqrt{2}} - \frac{\sqrt{3}}{2} Y_2 \frac{v_2}{\sqrt{2}}, 
	\hspace{1.1cm} M_E = Y_3 \frac{v_1}{\sqrt{2}} - \frac{\sqrt{3}}{2} Y_4 \frac{v_2}{\sqrt{2}}. \nonumber
\end{align}
In our convention, the fermionic mass matrices are diagonalized as follows:
\begin{align}
	U^T M_U U_c &= M_U^{\rm diag}, \\
	D^T M_D D_c &= M_D^{\rm diag}, \\
	E^T M_E E_c &= M_E^{\rm diag}.
\end{align}
An important aspect is that quark-lepton unification allows us to write the Yukawa matrices in terms of these mass matrices, and the scalar sector becomes more predictive than a generic 2HDM as we shall see below.

The scalar potential for $H_1$ and $H_2$ with quantum numbers $(\mathbf{1}, \mathbf{2}, 1/2)$ can be written as,
\begin{align}
	V(H_1, H_2) &= m_{11}^2 H_1^\dagger H_1 + m_{22}^2 H_2^\dagger H_2 - m_{12}^2 \left[\left( H_1^\dagger H_2 \right) + {\rm h.c.} \right] \nonumber \\ 
	&+ \frac{\lambda_1}{2} \left( H_1^\dagger H_1\right)^2 + \frac{\lambda_2}{2} \left( H_2^\dagger H_2 \right)^2 + \lambda_3 \left( H_1^\dagger H_1 \right) \left( H_2^\dagger H_2 \right) + \lambda_4 \left( H_1^\dagger H_2 \right) \left( H_2^\dagger H_1 \right) \nonumber \\
	&+ \left[ \frac{\lambda_5}{2} \left( H_1^\dagger H_2\right)^2 + \lambda_6 \left( H_1^\dagger H_1 \right) \left( H_1^\dagger H_2 \right) + \lambda_7 \left( H_2^\dagger H_2\right) \left( H_1^\dagger H_2 \right) + {\rm h.c.}\right].
\end{align}
The physical Higgs fields are defined by:
\begin{align}
\begin{pmatrix} H \\ h \end{pmatrix} & = \begin{pmatrix} \cos \alpha & \sin \alpha \\ -\sin \alpha & \cos \alpha \end{pmatrix}
\begin{pmatrix} H_1^0 \\ H_2^0 \end{pmatrix}, \\[1ex]
\begin{pmatrix} G \\ A \end{pmatrix} & = \begin{pmatrix} \cos \beta & \sin \beta \\ -\sin \beta & \cos \beta \end{pmatrix}
\begin{pmatrix} A_1^0 \\ A_2^0 \end{pmatrix}, \\[1ex]
\begin{pmatrix} G^\pm \\ H^\pm \end{pmatrix} & = \begin{pmatrix} \cos \beta & \sin \beta \\ -\sin \beta & \cos \beta \end{pmatrix}
\begin{pmatrix} H_1^\pm \\ H_2^\pm \end{pmatrix},
\end{align}
where $h$ is identified as the SM-like Higgs, $H$ is an additional neutral Higgs, $H_i^0, H_i^{\pm}, A^0_i$ are the neutral, charged and CP-odd components of the Higgs doublets, respectively, and $G, \,G^{\pm}$ are the would-be Goldstone bosons. 
The mixing angle $\beta$ is defined by the ratio of the vevs of the Higgs doublets, $\tan{\beta} = v_2/v_1$. The couplings of $h$ are SM-like in the alignment limit $\sin{(\beta - \alpha)} \to 1$, which corresponds to $\alpha = \beta - \pi/2$. The parameter $\cos(\beta-\alpha)$ can also be written as
\begin{align}
\cos^2{(\beta - \alpha)} = \frac{M_{L}^2 - M_h^2}{M_H^2 - M_h^2},
\label{eq:cosbetaalpha}
\end{align}
where the mass parameter $M_L$ is given by
\begin{align}
M_L^2 = v^2\left( \lambda_1 c_\beta^4 + \lambda_2 s_\beta^4 + 2\lambda_{345} s_\beta^2 c_\beta^2 + 2\lambda_6 c_\beta^2 s_{2\beta} + 2\lambda_7 s_\beta^2 s_{2\beta} \right),
\end{align}
where $v^2\!=\!v_1^2+v_2^2$ and from Eq.~\eqref{eq:cosbetaalpha} the decoupling limit becomes evident in the limit $M_H\!\gg\!M_L, M_h$. 

In the decoupling limit, the physical Higgs masses are given by
\begin{align}
	M_H^2 &= \frac{m_{12}^2}{s_{\beta}c_\beta} + v^2 \left[ \lambda_1 c_\beta^2 s_\beta^2 + \lambda_2 c_\beta^2 s_\beta^2 - 2\lambda_{345} c_\beta^2 s_\beta^2 - 2\lambda_6(ct_{\beta} + c_\beta s_\beta c_{2\beta}) + 2\lambda_7(t_{\beta} + c_\beta s_\beta c_{2\beta})\right], 
\end{align}
\begin{align}
	M_h^2 &= v^2\left( \lambda_1 c_\beta^4 + \lambda_2 s_\beta^4 + 2\lambda_{345} s_\beta^2 c_\beta^2 + 2\lambda_6 c_\beta^2 s_{2\beta} + 2\lambda_7 s_\beta^2 s_{2\beta} \right), \\
	M_A^2 &= \frac{m_{12}^2}{s_\beta c_\beta} - \frac{v^2}{2} (2\lambda_5 + \lambda_6 ct_{\beta} + \lambda_7 t_{\beta}),\\
	M_{H^{\pm}}^2 &= \frac{m_{12}^2}{s_\beta c_\beta} - \frac{v^2}{2} ( \lambda_4 + \lambda_5 + \lambda_6 ct_{\beta} + \lambda_7 t_{\beta}),
\end{align}
where $\lambda_{345}=\lambda_3+\lambda_4+\lambda_5$. Here $c\beta=\cos \beta$, $s\beta=\sin \beta$, $ct\beta=\cot \beta$ and $t\beta=\tan \beta$.
See Appendix~\ref{app:higgsmasses} for more details on the scalar potential and the masses of the scalar fields.
%
\section{HIGGS BOSONS DECAYS}
\label{sec:decays}
In this section, we discuss the decay properties of the new Higgs bosons and relations among the decay widths that arise from quark-lepton unification. For simplicity we assume the Yukawa interactions to be flavor-diagonal~\cite{Pich:2009sp}, this will be justified in Section~\ref{sec:flavor} where we discuss constraints from flavor violation. In the limit when $h$ is the SM-Higgs, $H$ does not interact with SM gauge bosons, so the following decays channels vanish at tree-level
\begin{align}
	\Gamma(H \to W^+ W^-) &= \Gamma(H \to ZZ) = 0, \\
	\Gamma(A \to W^+ W^-) &= \Gamma(A \to ZZ) = 0, \,\,\,
	\Gamma(H^{\pm} \to W^{\pm} Z) = 0.
\end{align}
Consequently, the total decay width of the heavy Higgs $H$ corresponds to 
\begin{align}
	\Gamma_{\rm T} (H) = \Gamma(H \to \bar{d}_i d_i) + \Gamma(H \to \bar{e}_i e_i) + \Gamma(H \to \bar{u}_i u_i) + \Gamma(H \to hh),
\end{align}
where the repeated index implies a sum over the different flavors. The trilinear coupling between $H$ and two SM Higgs bosons can be written as
\beq
\lambda_{\rm eff} = \frac{3}{2} \left[ \lambda_1s_{2\beta}(c_{2\beta}+1) + \lambda_2s_{2\beta}(c_{2\beta}-1) - 2 \lambda_{345} c_{2\beta}s_{2\beta}  - 2 \lambda_6(c_{2\beta} + c_{4\beta}) - 2\lambda_7(c_{2\beta} - c_{4\beta}) \right],
\eeq
we refer the reader to Appendix~\ref{sec:appFR} for a complete list of the Feynman rules.
\begin{figure}[t]
\centering
\includegraphics[width=0.49\linewidth]{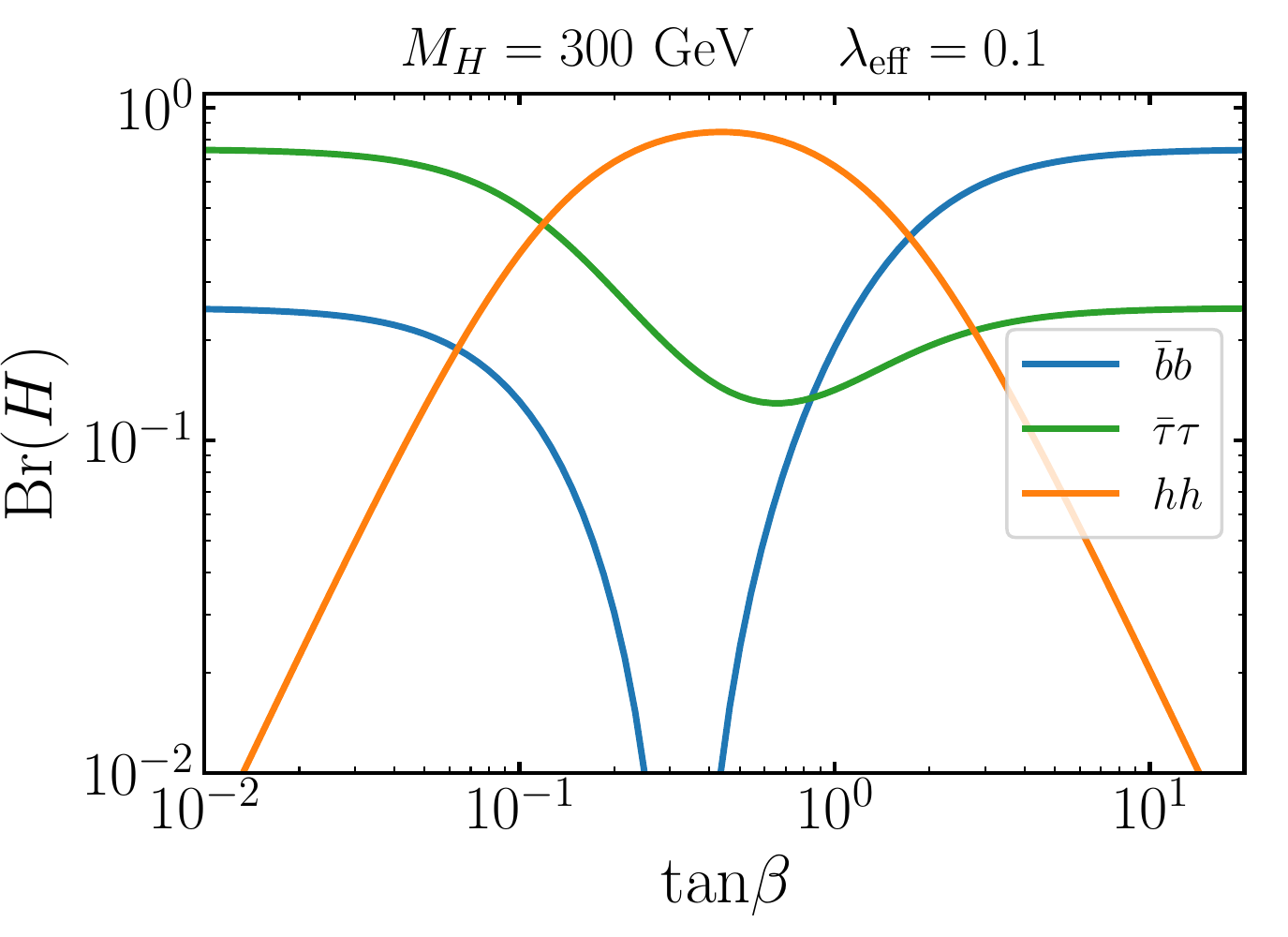}
\includegraphics[width=0.49\linewidth]{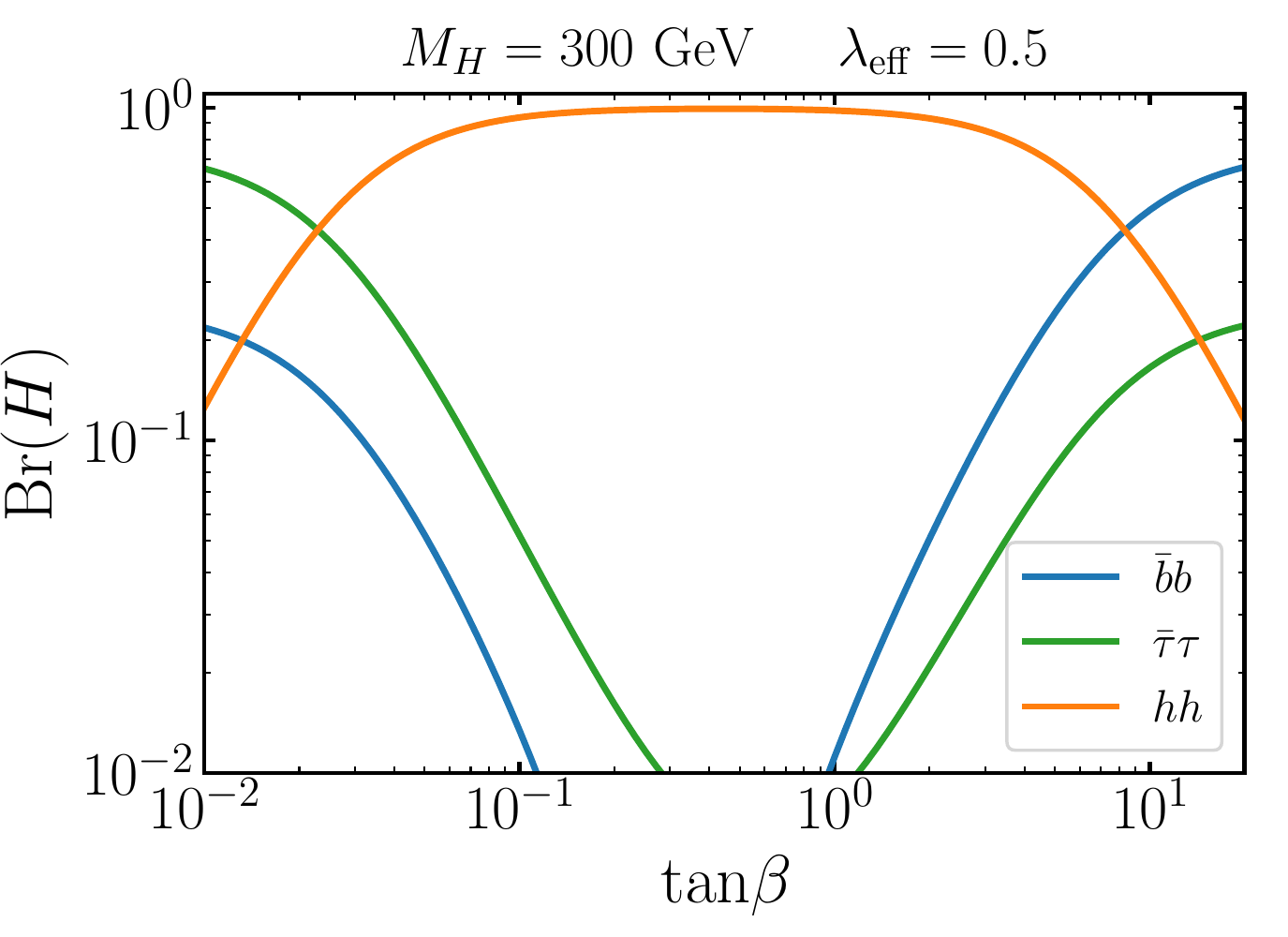}
\includegraphics[width=0.49\linewidth]{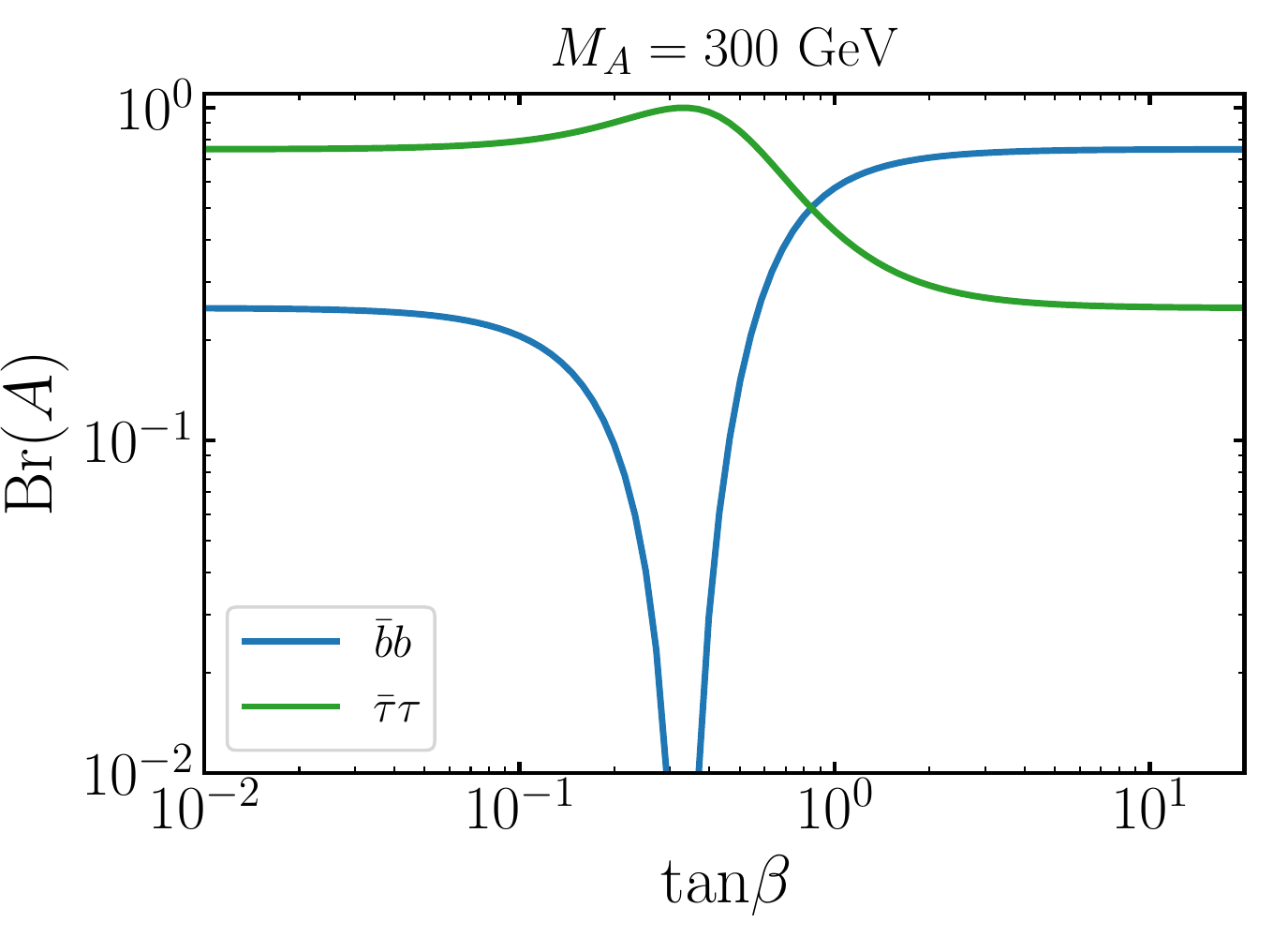}
\caption{Branching ratio for the different decay channels of $H$ and $A$ as a function of the parameter $\tan \beta$. In the upper panel, the two different plots corresponds to different values for the $H\!-\!h\!-\!h$ coupling $\lambda_{\rm eff}$ and we fix $M_{H,A} = 300 \ {\rm GeV}$.}
\label{fig:Hlowmass}
\end{figure}
In the limit with flavor-diagonal couplings this theory gives clean predictions for the coupling of $H$ and $A$ to down-type quarks and charged leptons. Namely, both couplings depend on the physical masses and the value of $\tan \beta$ as given in Appendix~\ref{sec:appFR}.

In the top panel in Fig.~\ref{fig:Hlowmass} we present our results for the branching ratio of $H$ as a function of $\tan \beta$. In this case we fix $M_H=300$ GeV so the decay $H \to \bar{t} t$ is kinematically closed. The blue (green) line shows the branching ratio for the decay channel $H  \to \bar{b} b$ ($H  \to \bar{\tau} \tau$). The orange line corresponds to the channel $H  \to h h$ which depends on the value of $\lambda_{\rm eff}$. As can be seen, the branching ratio for the $H  \to \bar{b} b$ channel nearly vanishes for $\tan\beta\approx 0.3$, this is because there is a cancellation between the two terms in the coupling to down-type quarks. 

For small values of $\tan\beta$, quark-lepton unification predicts the following relation
\beq
\Gamma(H \to \bar{\tau} \tau) = 3 \, \Gamma(H \to \bar{b} b),
\eeq
from the top-left panel in Fig.~\ref{fig:Hlowmass} we can see that this relation is already satisfied for $\tan \beta \lesssim 0.05$. For large values of $\tan \beta$  we have that
\beq
\Gamma(H \to \bar{\tau} \tau) = \frac{1}{3} \, \Gamma(H \to \bar{b} b),
\eeq
from the plot we can see that for $\tan \beta \gtrsim 3$ this relation is already satisfied.

In the lower panel in Fig.~\ref{fig:Hlowmass} we present our results for the branching ratio of $A$ as a function of $\tan \beta$.  In contrast to $H$, the pseudoscalar $A$ has no trilinear term with $hh$, this implies that the decay channel $A\to hh$ vanishes at tree-level. Quark-lepton unification gives the following relation for small values of $\tan \beta$
\beq
\label{eq:Asmall}
\Gamma(A \to \bar{\tau} \tau) = 3 \, \Gamma(A \to \bar{b} b),
\eeq
and for large values of $\tan \beta$
\beq
\label{eq:Alarge}
\Gamma(A \to \bar{\tau} \tau) = \frac{1}{3} \, \Gamma(A \to \bar{b} b),
\eeq
from this plot we can see that Eq.~\eqref{eq:Asmall} is already satisfied for $\tan \beta \lesssim 0.05$ while Eq.~\eqref{eq:Alarge} is already satisfied for $\tan \beta \gtrsim 3$.
\begin{figure}[t]
\centering
\includegraphics[width=0.49\linewidth]{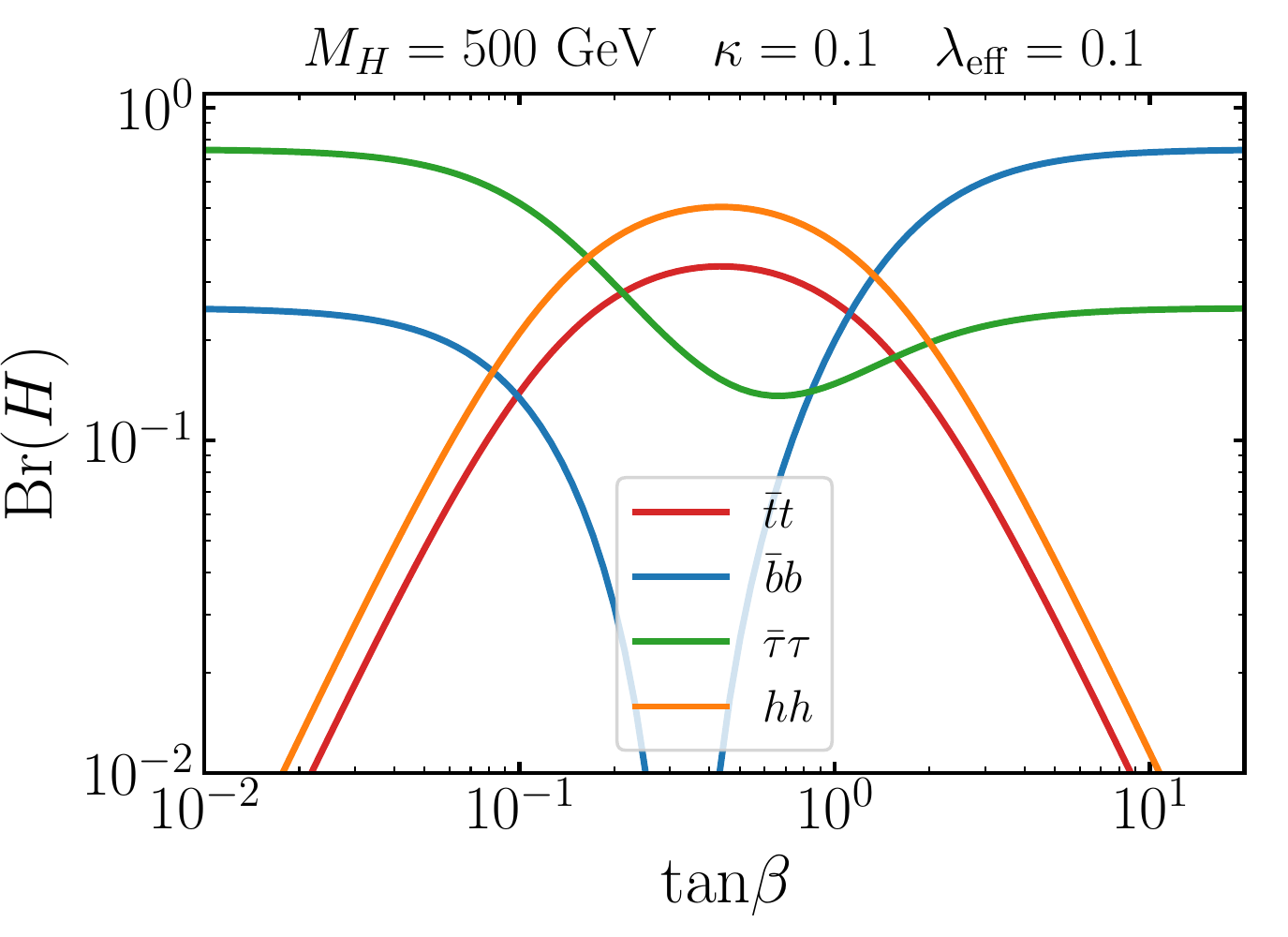}
\includegraphics[width=0.49\linewidth]{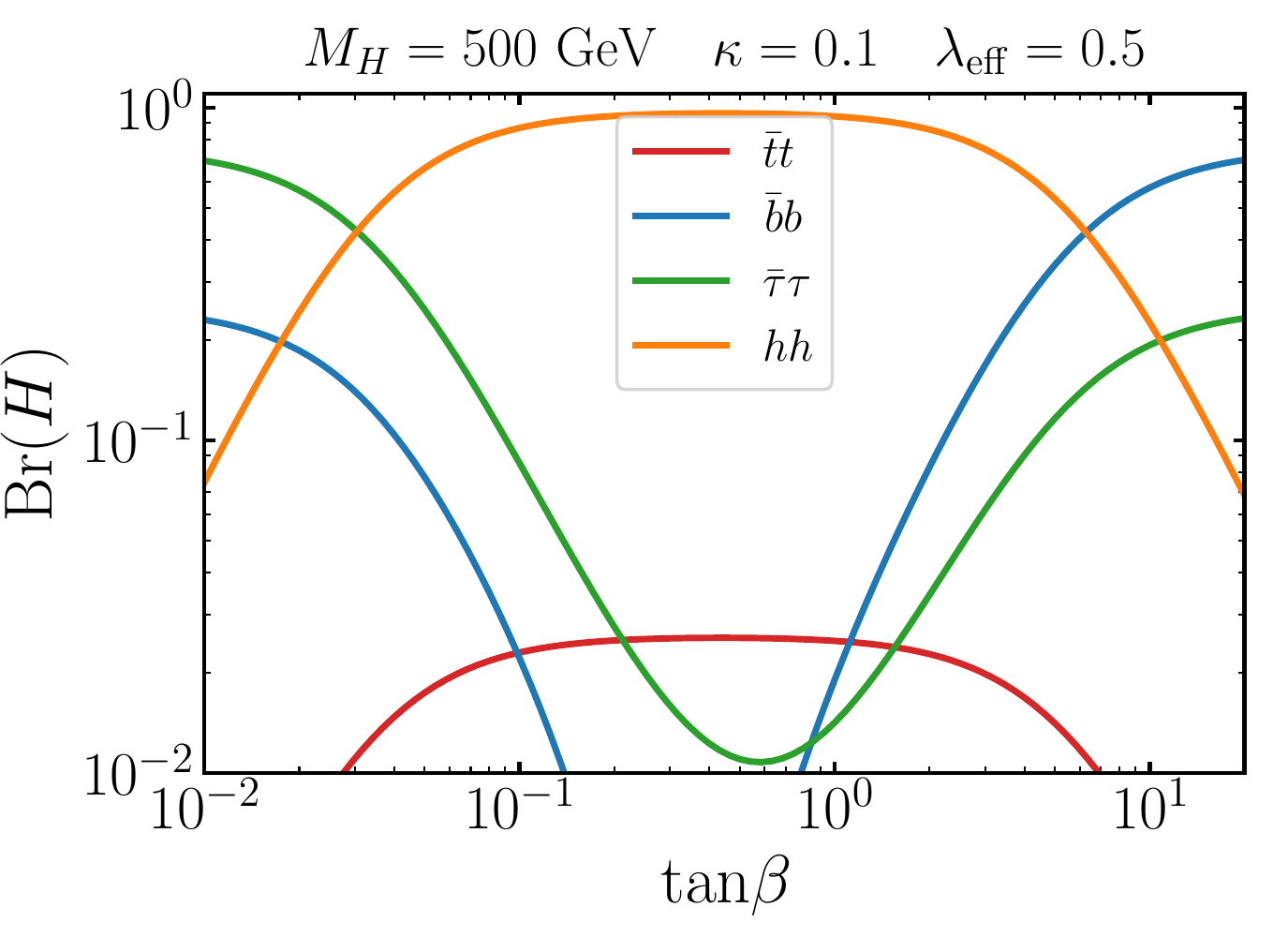}
\includegraphics[width=0.49\linewidth]{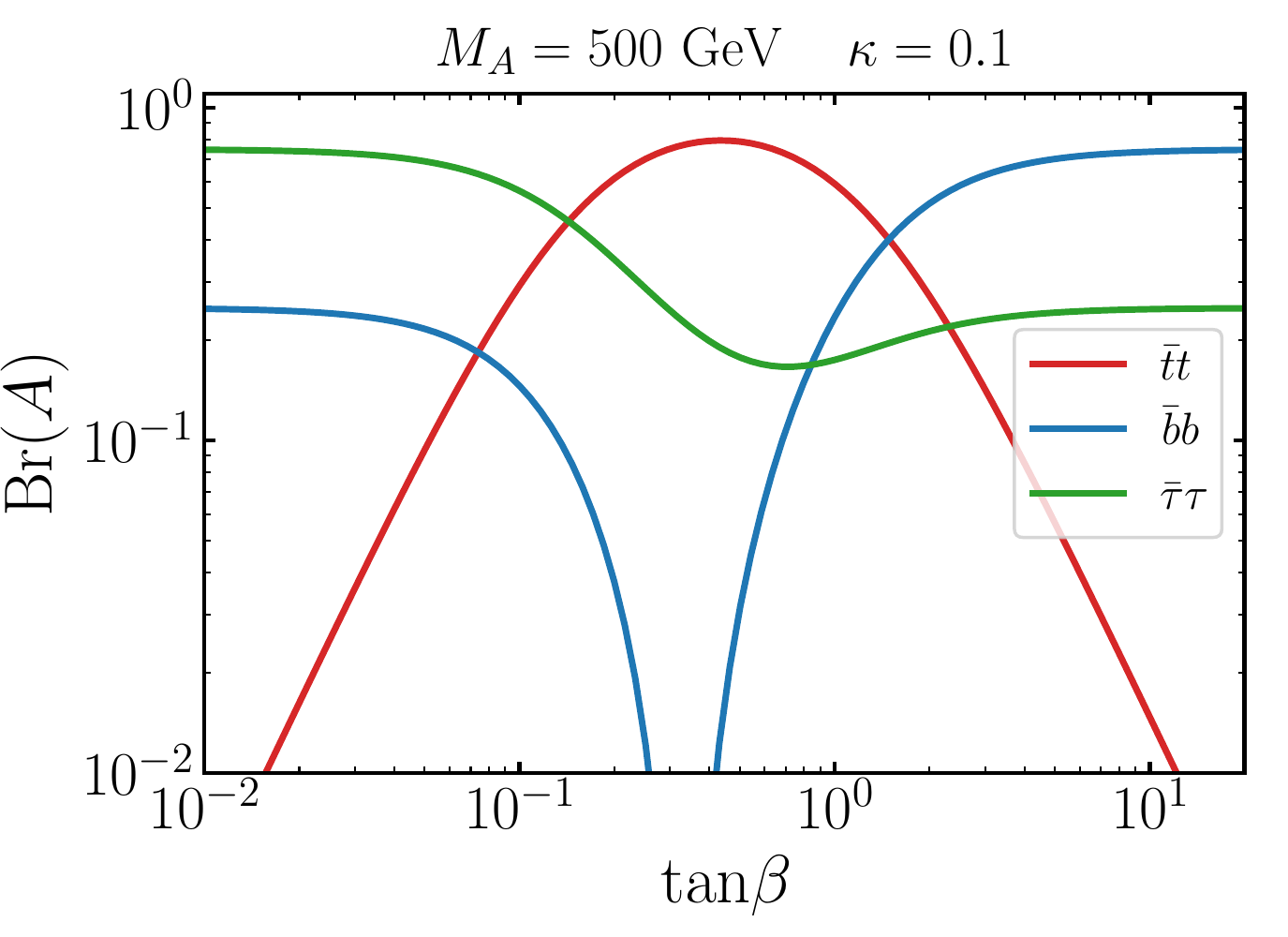}
\includegraphics[width=0.49\linewidth]{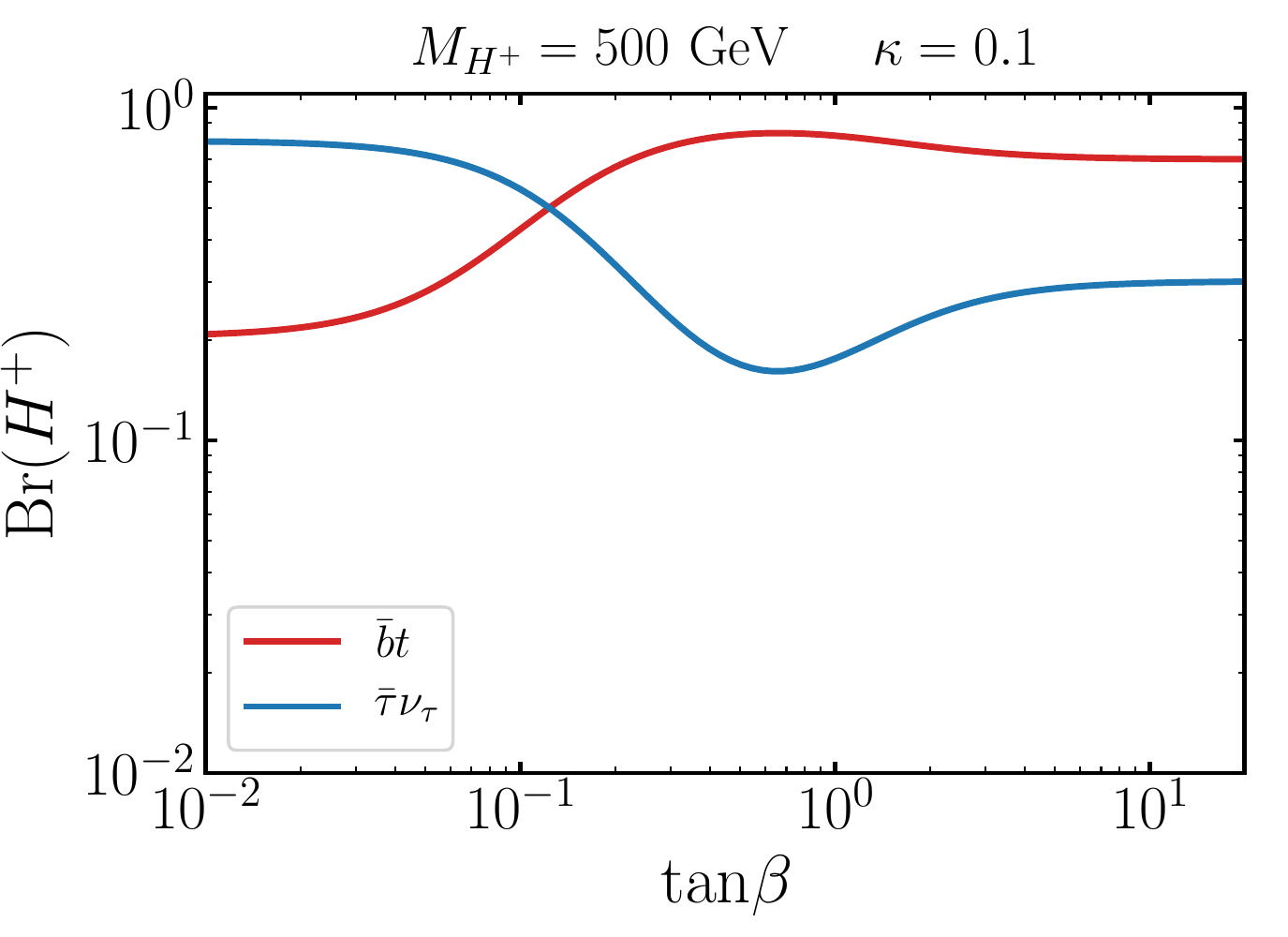}
\caption{\textit{Upper panel:} Branching ratio for the different decay channels of the scalar $H$ as a function of the parameter $\tan \beta$. Different plots correspond to different values for the $H\!-\!h\!-\!h$ coupling $\lambda_{\rm eff}$ and we fix $\kappa=0.1$ and $M_H=500$ GeV. \textit{Lower panel:} On the left (right) panel we show the branching ratio for the different decay channels of the pseudoscalar $A$ (charged Higgs $H^{\pm}$) as a function of the parameter $\tan \beta$ and we set $M_A=M_{H^\pm}=500$ GeV.}
\label{fig:Hhighmass}
\end{figure}

Unfortunately, the theory does not predict the coupling to up-type quarks. However, we can parametrize this coupling by introducing the parameter $\kappa$,
\beq
\label{eq:kappa}
C^H_{uu} = C^A_{uu} = \frac{1}{\sqrt{2}} C_{Lud} = \frac{\kappa}{4 v} M_U^{\rm diag},
\eeq
where $C_{Lud}$ corresponds to the coupling with the charged scalar. Since there is freedom in the $U_C^T M_\nu^{D \,T} U$ term, it can be fixed at each point in order to remove the dependence on the parameter $\tan \beta$. Since the branching ratios shown in Fig.~\ref{fig:Hlowmass} are independent of the parameter $\kappa$, the LHC bounds can be avoided by choosing a small value for this parameter.

In Fig.~\ref{fig:Hhighmass} we show the branching ratios for $M_{H,A}=500$ GeV when the decay $H,A \to \bar{t}t$ is kinematically open. In the bottom-right panel we show the branching ratios for the decay channels of the charged Higgs as a function of $\tan \beta$. The decay width for $H^+\to \bar{b} t$ depends on the parameter $\kappa$ given in Eq.~\eqref{eq:kappa}, which we fix as before to $\kappa=0.1$ in both plots; however this decay width changes as we vary $\kappa$. Since the right-handed neutrinos acquire their mass from the $\SU(4)_C$ symmetry breaking scale they are expected to be heavy, and hence, the decay channel $H^+\to \bar{e}_i N_j$ is kinematically closed.

As we discussed in Ref.~\cite{Perez:2021mgz}, the idea of quark-lepton unification also predicts relations among the decay widths of the scalar leptoquarks present in the theory. Assuming flavor-diagonal Yukawa interactions, we obtain the following relations between the decay widths of the leptoquarks
\begin{align}
\frac{\Gamma(\phi_3^{-2/3} \to \bar{b} \tau)}{M_{\phi_3^{-2/3}}} = \frac{\Gamma(\phi_4^{-2/3} \to \bar{b} \tau)}{M_{\phi_4^{-2/3}}} =\frac{\Gamma(\phi_4^{5/3} \to \bar{\tau} t)}{M_{\phi_4^{5/3}}} = \frac{\Gamma(\phi_3^{1/3} \to \bar{b} \nu_\tau)}{M_{\phi_3^{-2/3}}} = \frac{9 ({M_b + M_\tau)^2}}{32 \pi v^2 \sin^2{\beta}} .
\end{align}
Consequently, if scalar leptoquarks are discovered in the near future, then these relations can be used to test whether the underlying theory comes from quark-lepton unification. A detailed study of the collider phenomenology for the scalar leptoquarks is beyond the scope of this paper.
%
%
\section{PRODUCTION AT THE LHC}
\label{sec:LHC}
%
\begin{figure}[h]
\centering
\includegraphics[scale=0.75]{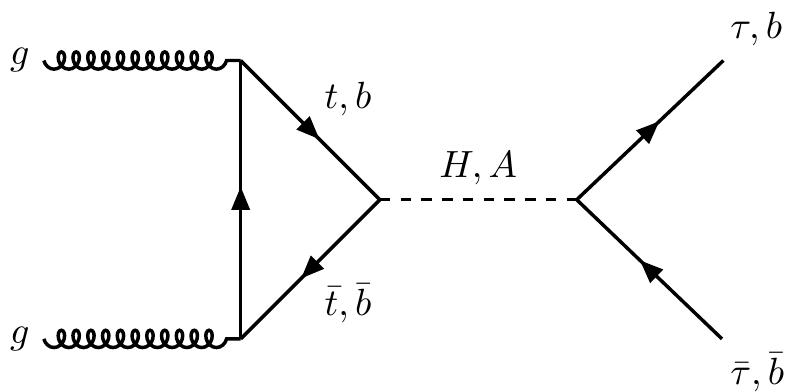}
\caption{Feynman diagram for the production of the new scalars $H$ and $A$ via gluon fusion.}
\label{fig:hgg}
\end{figure}
The new neutral Higgs bosons can be produced at the LHC through gluon fusion with the top and the bottom quarks running in the loop as shown in the Feynman diagram in Fig.~\ref{fig:hgg}. The collider phenomenology in the 2HDM has been studied before in different contexts, see e.g. \cite{Chiang:2013ixa,Hespel:2014sla,BhupalDev:2014bir,Wang:2017xml,Chowdhury:2017aav,Grzadkowski:2018ohf,Chen:2019pkq,Kling:2020hmi,Arco:2020ucn,Kanemura:2021dez,Wang:2022yhm}. The contribution from the bottom quark is relevant only for large and small values of $\tan\beta$, and hence, the cross-section mostly depends on the $\kappa$ parameter used to parametrize the coupling to the top-quark. The effective coupling between the neutral Higgs bosons and the gluons is 
given by 
\beq
\mathcal{L} \supset g_{ggH} \frac{H}{v} G^{\mu\nu} G_{\mu\nu} +  g_{ggA} \frac{A}{v} G^{\mu\nu} \widetilde{G}_{\mu\nu},
\eeq
where the dual field strength tensor is given by $\widetilde{G}_{\mu\nu} = \varepsilon_{\mu\nu\alpha\beta}G^{\alpha\beta}/2$ and 
\begin{align}
g_{ggH} & = \frac{\alpha_s}{8 \pi} \sum_q \tau_q [1+(1-\tau_q)f(\tau_q)] C^H_{qq} \frac{v}{M_q}, \\[1ex]
g_{ggA} & = \frac{\alpha_s}{8 \pi} \sum_q \tau_q f(\tau_q) C^A_{qq} \frac{v}{M_q},
\end{align}
where the sum is over the quarks in the SM, although the dominant contribution comes from the top quark, and the $f(\tau)$ loop function is given by
\begin{equation}
 f(\tau) =
    \begin{cases}
      \arcsin^2 \left( \sqrt{\tau^{-1}} \right)  & \tau \geq 1\\[1ex]
-\displaystyle \frac{1}{4} \left( \log \frac{ 1+ \sqrt{ 1 - \tau} }  {1-\sqrt{1-\tau}} - i \pi \right)^2 & \tau < 1
    \end{cases}       
\end{equation}
where $\tau \equiv 4M_q^2/M_H^2$. We are interested in the regions with large or small values of $\tan \beta$, since in these regions by measuring $pp\to H,A\to\bar{\tau}\tau$ then the cross-section for $pp\to H,A\to\bar{b}b$ can be predicted by using Eqs.~\eqref{eq:Asmall} and~\eqref{eq:Alarge}. We implement the model using \texttt{FeynRules 2.0}~\cite{Alloul:2013bka} and calculate the cross-sections using \texttt{MadGraph5\_aMC@NLO}~\cite{Alwall:2014hca} which were cross-checked in \texttt{Mathematica} with use of the MSTW2008~\cite{Martin:2009iq} set of parton distribution functions. 

\begin{figure}[t]
\centering
\includegraphics[width=0.495\linewidth]{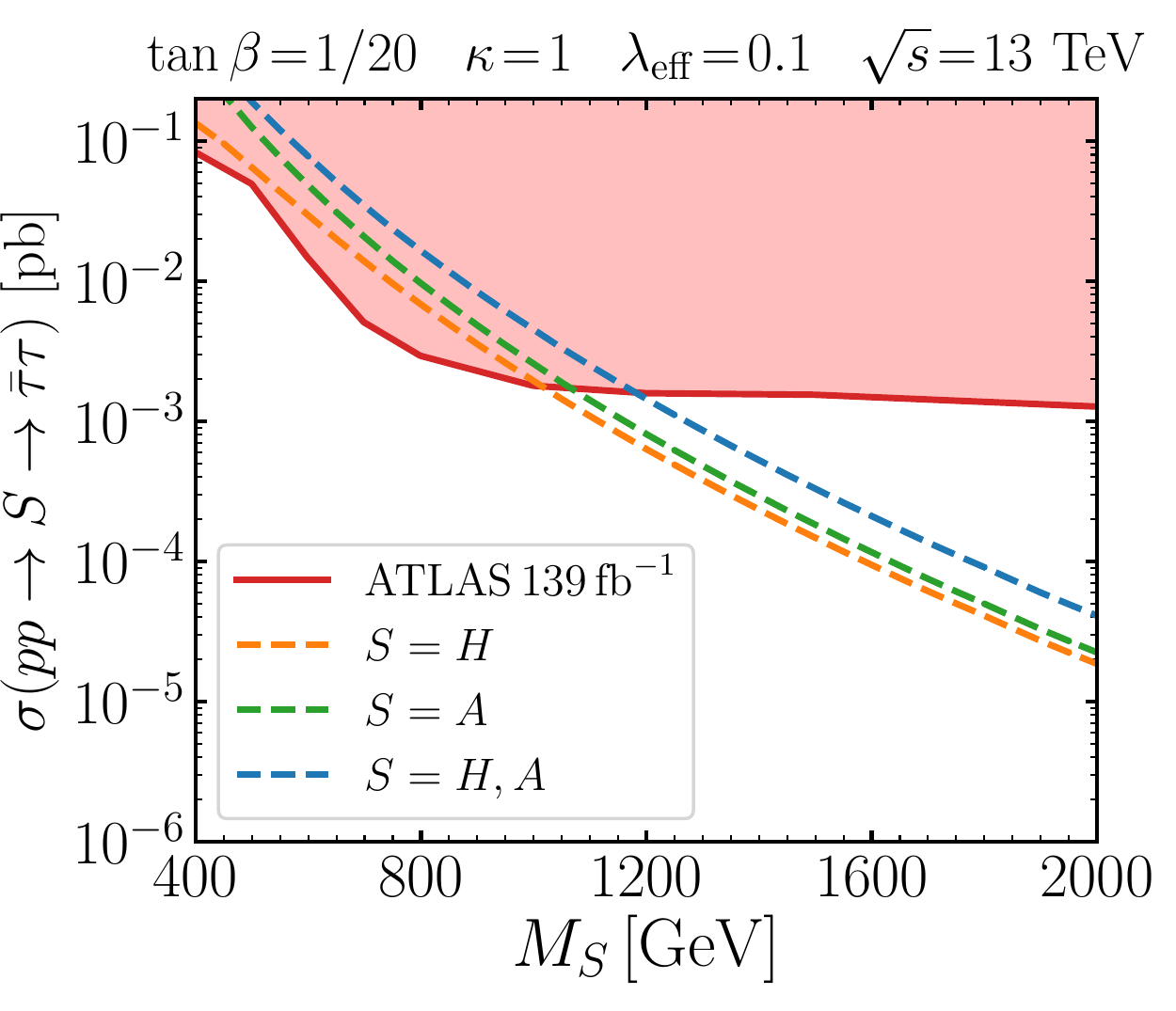}
\includegraphics[width=0.495\linewidth]{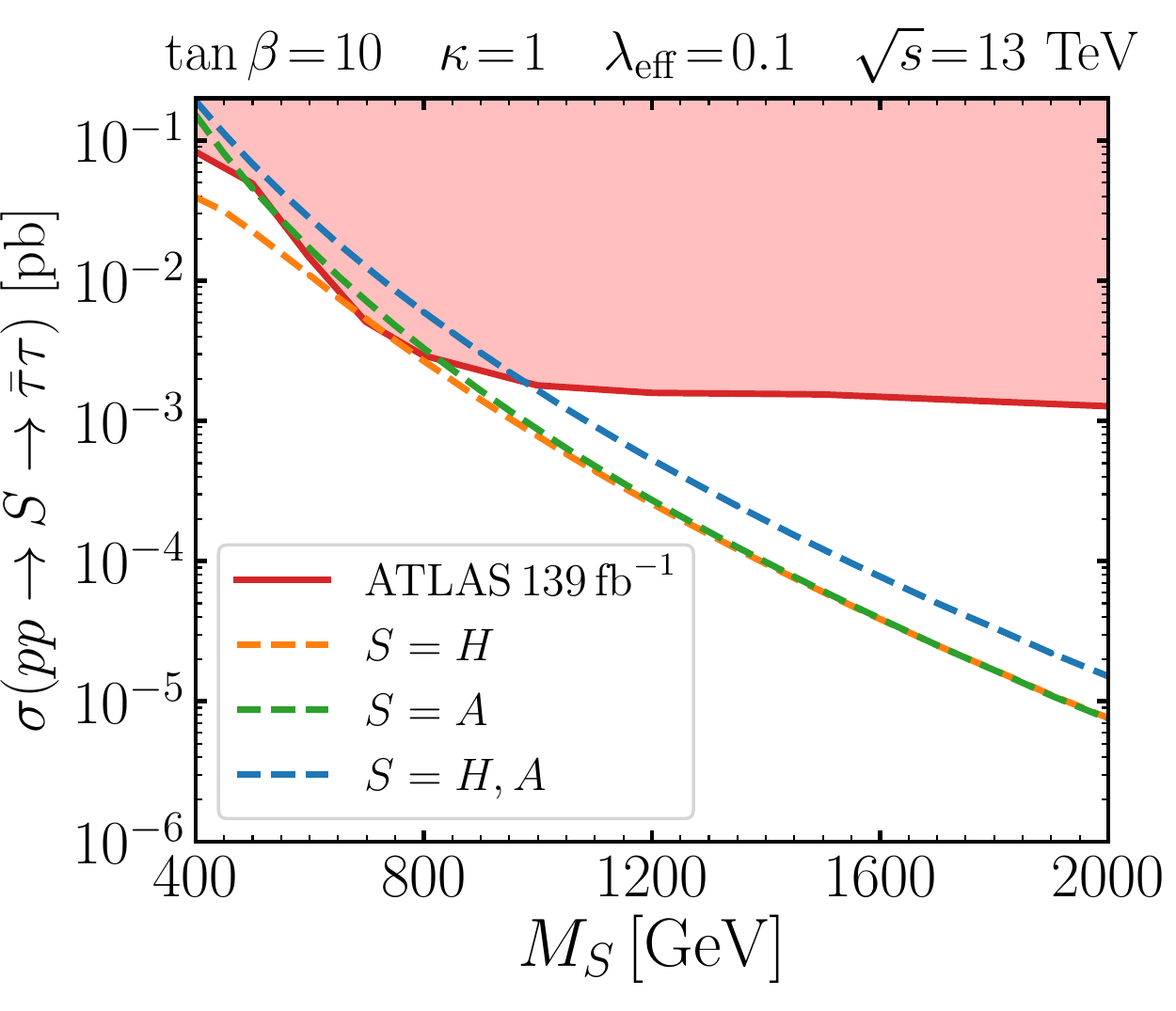}
\includegraphics[width=0.495\linewidth]{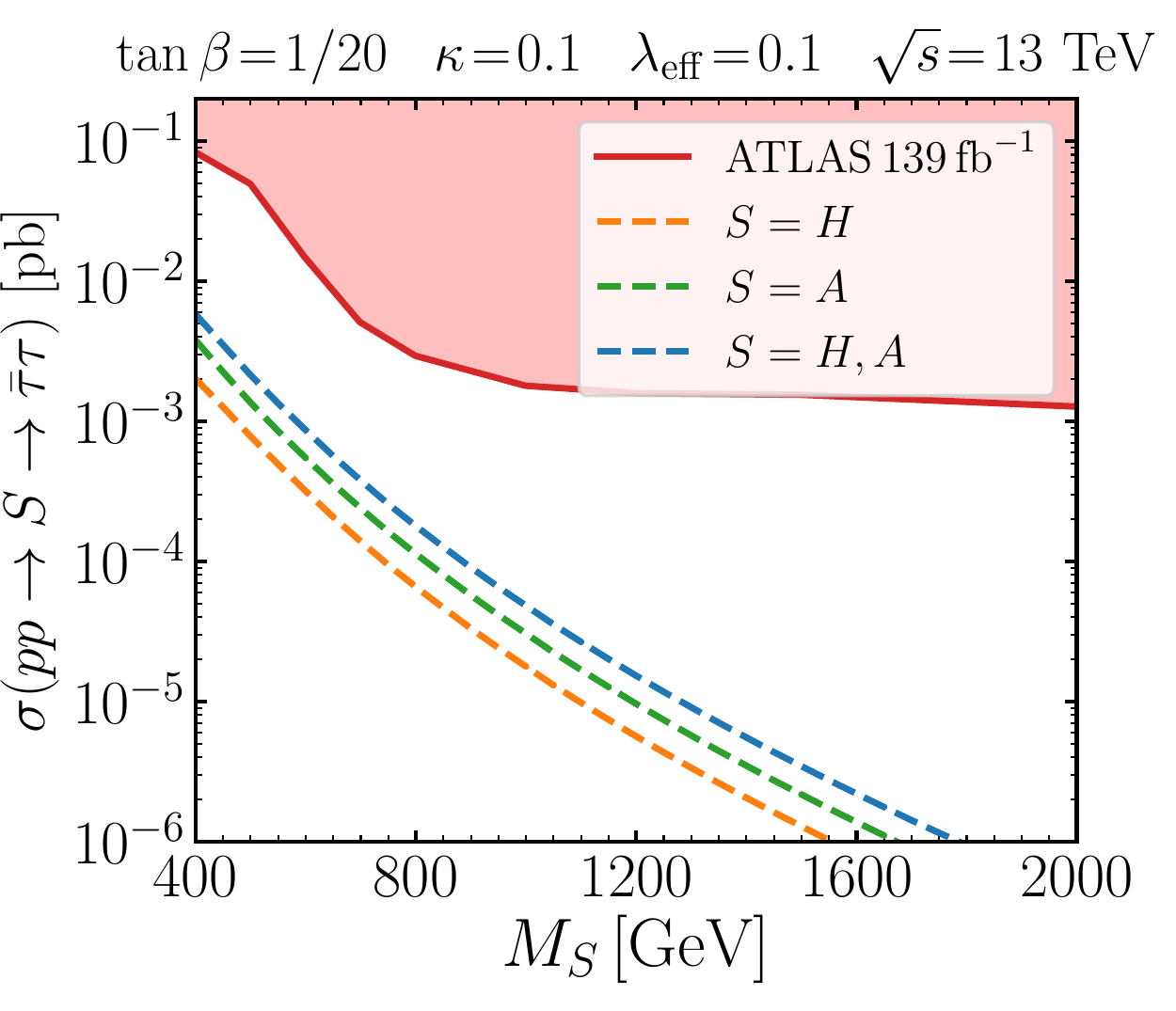}
\includegraphics[width=0.495\linewidth]{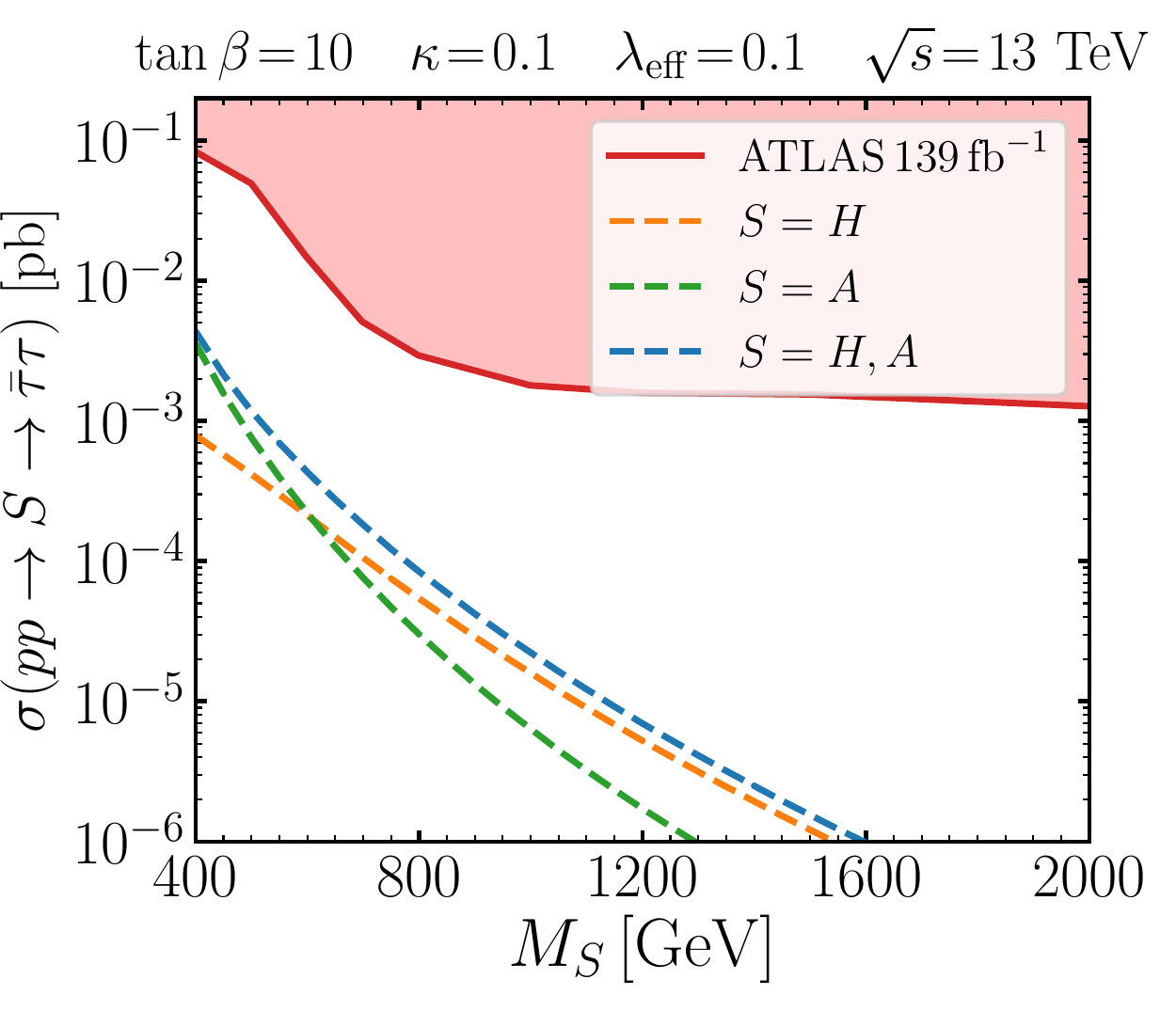}
\caption{Production cross-section for the process $pp \to S \to \bar{\tau} \tau$. The region shaded in red corresponds to the experimental limit by ATLAS~\cite{ATLAS:2020zms}. The dashed orange (green) line corresponds to the production of $H$ ($A$), while the dashed blue line takes into account both.  In the top-left panel we fix $\tan\beta\!=\!1/20$ and $\kappa=1$, in the top-right panel $\tan\beta\!=\!1$ and $\kappa=1$, in the bottom-left panel we fix $\tan\beta\!=\!1/20$ and $\kappa=0.1$, and in the bottom-right panel $\tan\beta\!=\!1$ and $\kappa=0.1$. In the four cases we fix $\lambda_{\rm eff} = 0.1$.}
\label{fig:tautau}
\end{figure}

In Fig.~\ref{fig:tautau} we present our predictions for the cross-section $pp\to S \to\bar{\tau}\tau$ as a function of the mass of the scalar $M_S$ with center-of-mass energy of $\sqrt{s}=13$ TeV. We implement the following cuts on the transverse momentum and the rapidity of the tau leptons, $p_T>30$ GeV and $|\eta|<2.5$. The dashed orange line corresponds to the case $S=H$, the green dashed line is for $S=A$ and the dashed blue line includes the contribution from both $S=H,A$ and assumes $M_H\!=\!M_A$, since for large masses their mass splitting cannot be large due the perturbativity of the scalar couplings and the constraints coming from electroweak precision observables~\cite{Haller:2018nnx}. The region shaded in red corresponds to the exclusion limit  from searches by the ATLAS~\cite{ATLAS:2020zms} collaboration for a heavy scalar decaying into a two tau leptons with integrated luminosity of 139 ${\rm fb}^{-1}$ (see also Ref.~\cite{Hou:2022nyh}). We fix the trilinear coupling to $\lambda_{\rm eff}=0.1$ which affects only the process involving $H$. 

The two upper panels in Fig.~\ref{fig:tautau} correspond to $\kappa=1$. For the plot on the left we set $\tan \beta=1/20$ and the LHC bound require the mass of $H$ and $A$ to be above 1.2 TeV; for the plot on the right we set $\tan \beta=10$ which requires the masses to be above 1 TeV. These bounds can be avoided by choosing a smaller value for $\kappa$. In the lower panels we set $\kappa=0.1$ and then the heavy scalars can be around the electroweak scale.

\begin{figure}[t]
\centering
\includegraphics[width=0.495\linewidth]{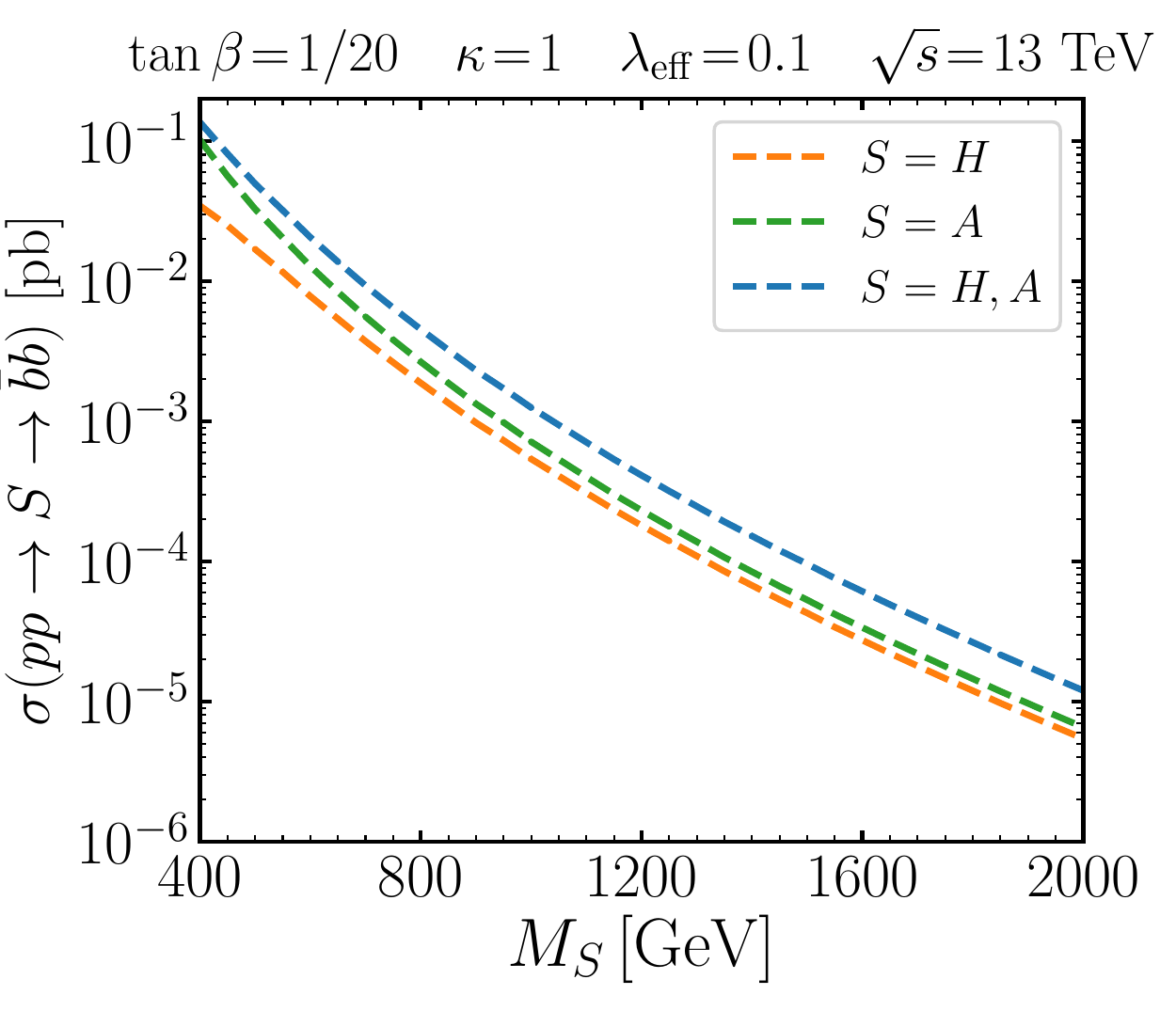}
\includegraphics[width=0.495\linewidth]{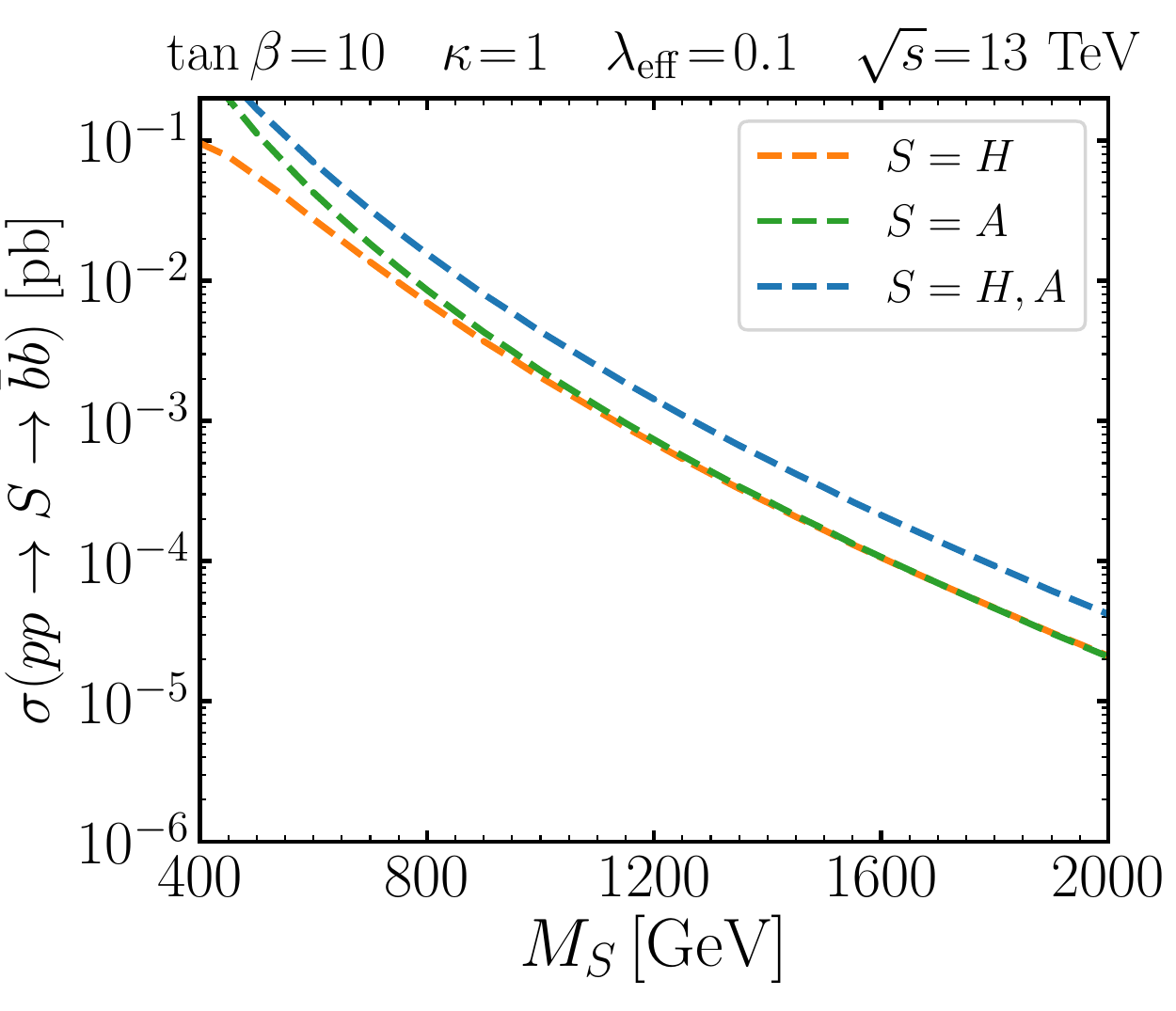}
\includegraphics[width=0.495\linewidth]{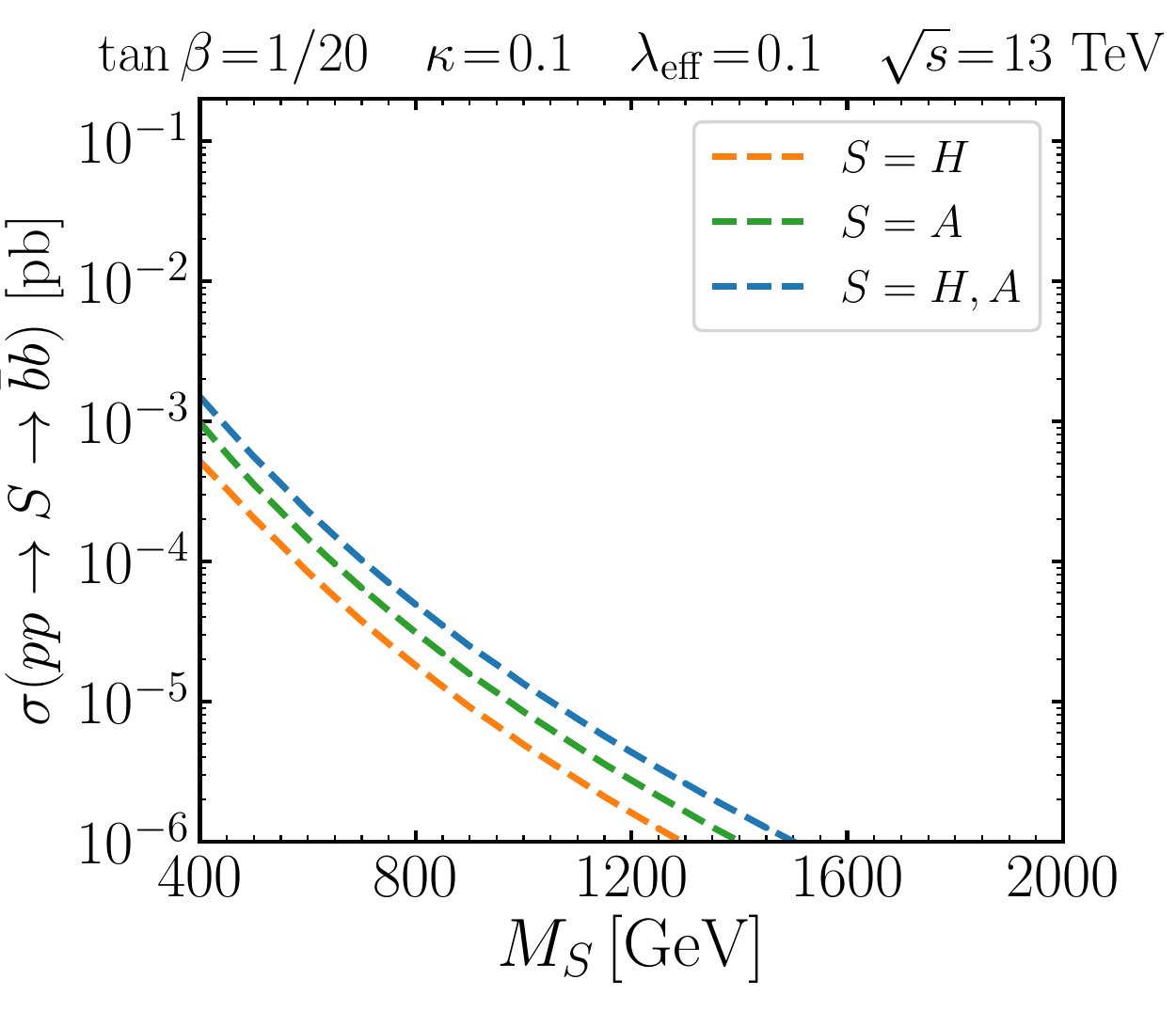}
\includegraphics[width=0.495\linewidth]{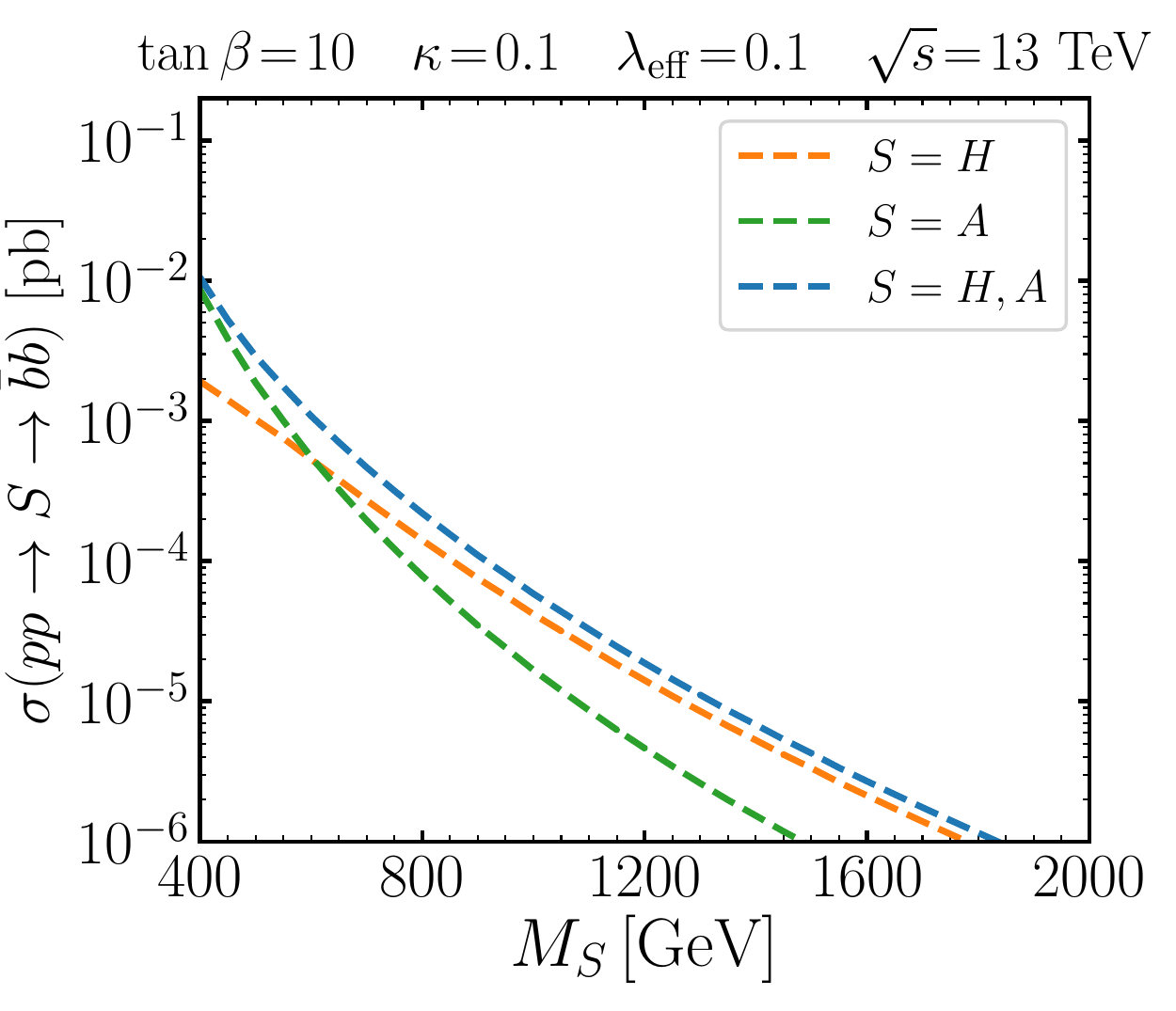}
\caption{Production cross-section for the process $pp \to S \to \bar{b} b$. The dashed orange (green) line corresponds to the production of $H$ ($A$), while the dashed blue line takes into account both.  In the top-left panel we fix $\tan\beta\!=\!1/20$ and $\kappa=1$, in the top-right panel $\tan\beta\!=\!1$ and $\kappa=1$, in the bottom-left panel we fix $\tan\beta\!=\!1/20$ and $\kappa=0.1$, and in the bottom-right panel $\tan\beta\!=\!1$ and $\kappa=0.1$. In the four cases we fix $\lambda_{\rm eff} = 0.1$.}
\label{fig:bb}
\end{figure}
In Fig.~\ref{fig:bb} we present the cross-sections for the process $pp\to S \to\bar{b}b$ with center-of-mass energy of $\sqrt{s}=13$ TeV for $\kappa=1$ ($\kappa=0.1$) in the upper (lower) panels. In this case we impose $p_T>50$ GeV for the transverse momentum and $|\eta|<2$ for the rapidity of the bottom quarks. These cross-sections have a similar magnitude as the ones with $\bar{\tau}\tau$ in the final state; however, the search for $\bar{b}b$ is more challenging experimentally and the current bounds are much weaker than the values predicted~\cite{CMS:2018kcg,CMS:2018hir}. Nonetheless, these predictions will be relevant for future searches of a heavy scalar decaying into two bottom quarks. We checked that for large and small values of $\tan \beta$ the relations from quark-lepton unification, Eqs.~\eqref{eq:Asmall} and~\eqref{eq:Alarge}, are satisfied.
 
For the charged Higgs the most relevant bound comes from the experimental measurement of the transition $b \to s \gamma$ which requires $M_{H^\pm}> 790$ GeV~\cite{Misiak:2020vlo,Atkinson:2021eox}; however, this observable depends on the coupling between the charged Higgs and the top quark which is not predicted in this theory and we parametrize in Eq.~\eqref{eq:kappa} using the parameter $\kappa$. That bound corresponds to setting $\kappa\simeq 1$ and it becomes weaker for smaller values of this parameter. Regarding production at the LHC, the charged Higgs can be pair-produced through a $Z$ boson or a photon; however, this cross-section is smaller than the one we have consider for single production of the neutral Higgs bosons, and hence, we do not discuss it any further. The high luminosity stage at the LHC is expected to reach an integrated luminosity of 3000 ${\rm fb}^{-1}$ and will probe masses for $H$ and $A$ around the TeV scale. Consequently, the relations between the decay widths predicted from quark-lepton unification can be tested in the near future.
\section{HIGGS FLAVOR VIOLATION}
\label{sec:flavor}
In the previous study we neglected the flavor-violating Higgs decays because they can be generically suppressed. There are many studies of flavor violation in the 2HDM see e.g. Refs.~\cite{Diaz:2000cm,Paradisi:2006jp,WahabElKaffas:2007xd,Deschamps:2009rh,Liu:2015oaa,Davidson:2016utf,Arnan:2017lxi,Altmannshofer:2018bch,Babu:2018uik,Vicente:2019ykr,Primulando:2019ydt,Ghosh:2020tfq,Atkinson:2022pcn}, here we study the constraints from flavor violation that arise in the context of quark-lepton unification.

The interactions between $H$ and the SM down-type quarks and charged leptons are defined by
\begin{align}
C_{dd}^H & = \left( 3\tan{\beta} - \cot{\beta}  \right) \frac{M_D^{\rm diag}}{4v} + \left( \tan{\beta} + \cot{\beta}\right)\frac{V_c^{*} M_E^{\rm diag} V^{\dagger} }{4 v}, \label{eq:cddH} \\[1.5ex]
C_{ee}^H & = \left( \tan{\beta} - 3\cot{\beta}  \right) \frac{M_E^{\rm diag}}{4v} + 3\left( \tan{\beta} + \cot{\beta}\right)\frac{V_c^{T} M_D^{\rm diag} V}{4 v}, \label{eq:ceeH}
%
%
%
\end{align}
where $V=D^\dagger E$ and $V_c=D_c^\dagger E_c$ have the information about the unknown mixings between the quarks and leptons. Notice that in the above equations, the first term is flavor-diagonal 
while the second term generically violates flavor but their values are bounded by the quarks or lepton masses. The framework of quark-lepton unification implies that the flavor violating couplings in the quark sector are proportional to the lepton masses and vice versa.

\begin{figure}[t]
\centering
\includegraphics[width=0.475\linewidth]{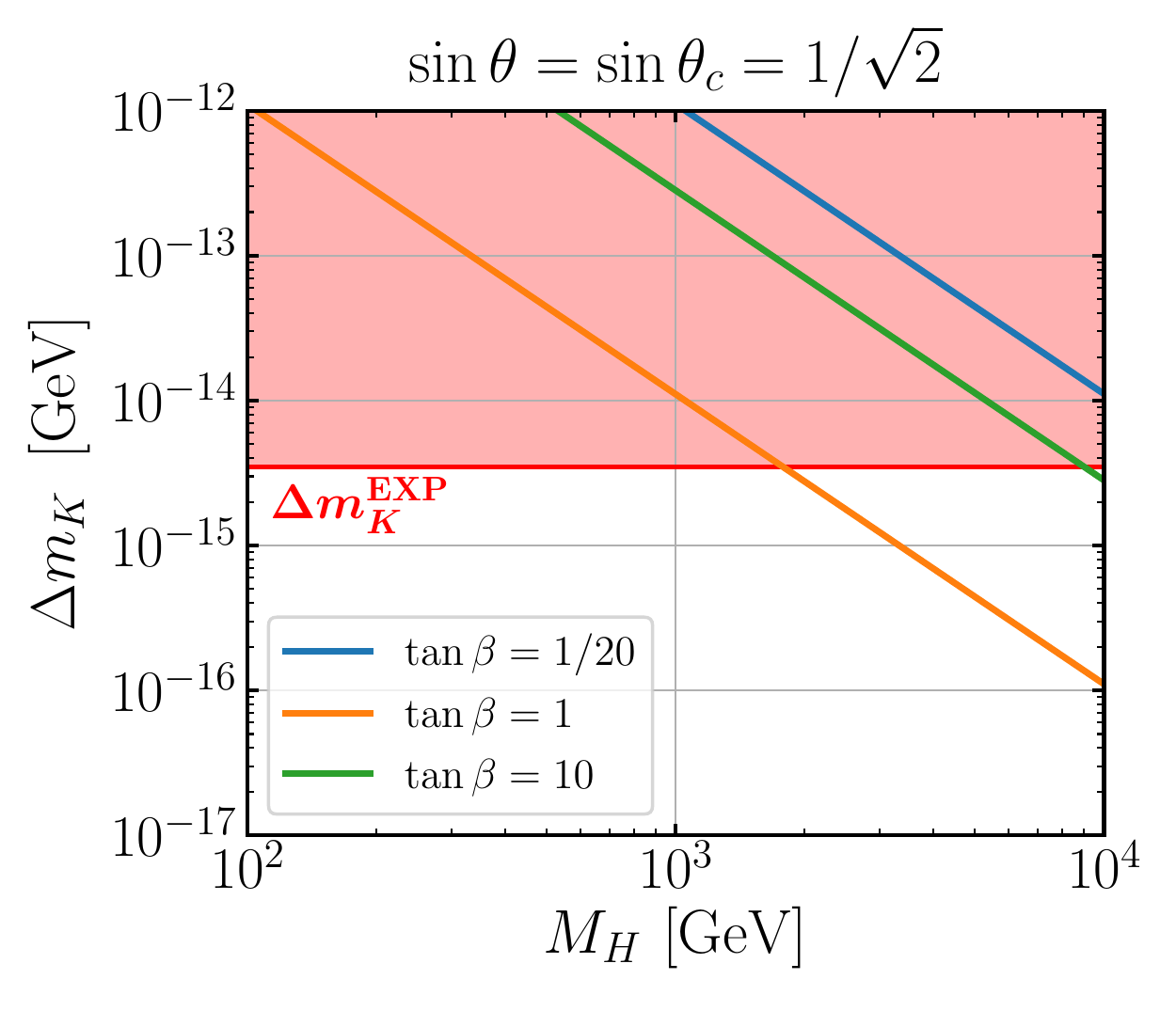}
\includegraphics[width=0.475\linewidth]{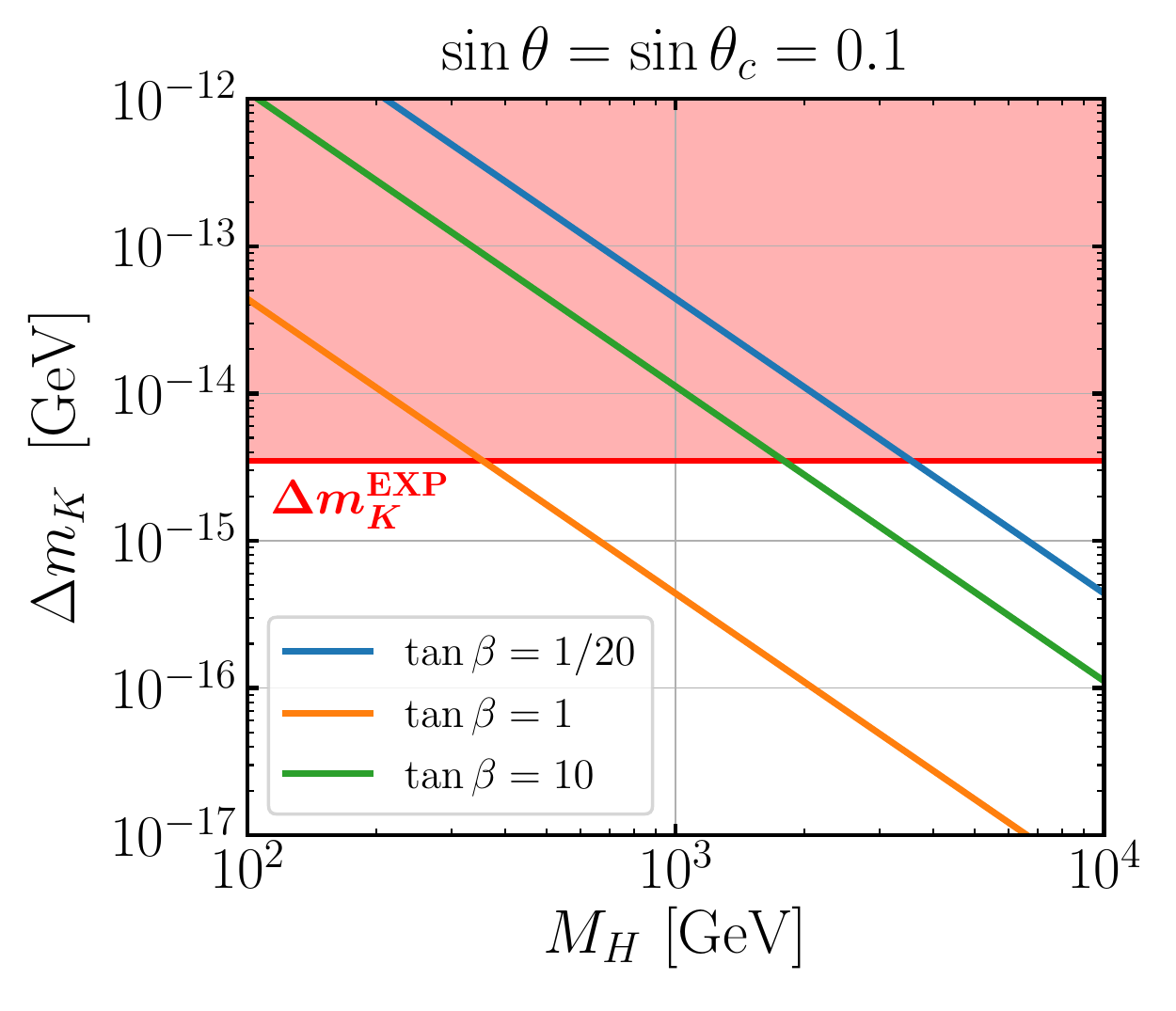}
\caption{Experimental constraint from $K^0-\bar{K}^0$  mixing on the parameter space. We show the results for $\Delta m_K$ in units of GeV as a function of the mass of the Higgs bosons $M_H=M_A$; the lines with different colors correspond to different values for $\tan \beta$ as shown in the legend. The region shaded in red gives a larger contribution than the measured value of $\Delta m_K^{\rm EXP}$. For the left (right) panel we fix the mixing angles to $\sin \theta= \sin \theta_c=1/\sqrt{2}$ ($\sin \theta= \sin \theta_c=0.1$).   }
\label{fig:kmixing}
\end{figure}

From Eqs.~\eqref{eq:cddH} and \eqref{eq:ceeH} we see that the effects of flavor violation are proportional to the fermion masses, and hence, the largest effect will involve the third generation. Here we discuss a simple scenario where all flavor violating processes are suppressed by the masses of the quarks and leptons from the first and second generations. In this scenario $V$ and $V_c$ are given by
\begin{equation}
V = D^\dagger E = \begin{pmatrix} \cos \theta && \sin \theta &&  0 \\ 
- \sin \theta && \cos \theta &&  0
\\
0 && 0 && 1  \end{pmatrix} \qquad \text{ and } 
\qquad 
V_c = D_c^\dagger E_c =\begin{pmatrix} \cos \theta_c && \sin \theta_c &&  0 \\ 
- \sin \theta_c && \cos \theta_c &&  0
\\
0 && 0 && 1  \end{pmatrix}.
\label{textures}
\end{equation}
Therefore, the elements $(C_{dd}^H)^{a3}=(C_{dd}^H)^{3a}=(C_{ee}^H)^{a3}=(C_{ee}^H)^{3a}=0$ and $(C_{dd}^H)^{33}=(C_{ee}^H)^{33}=1$ with $a=1,2$.
The interactions of the CP-odd Higgs, $A$, also can be written in a similar way,
\begin{align}
C_{dd}^A & = \left( \cot{\beta} - 3\tan{\beta}  \right) \frac{M_D^{\rm diag}}{4v} - \left( \tan{\beta} + \cot{\beta}\right)\frac{V_c^{*} M_E^{\rm diag} V^\dagger }{4 v}, \\[1.5ex]
C_{ee}^A & = \left( 3\cot{\beta} - \tan{\beta}  \right) \frac{M_E^{\rm diag}}{4v} - 3\left( \tan{\beta} + \cot{\beta}\right)\frac{V_c^{T} M_D^{\rm diag} V}{4 v}.
\label{eq:yaee}
\end{align}
The relations in Eqs.~\eqref{eq:cddH}-\eqref{eq:yaee} can be seen as an ansatz for the Yukawa matrices which is motivated by quark-lepton unification. When we use Eq.~(\ref{textures}) one can easily find that $(C_{dd}^A)^{a3}=(C_{dd}^A)^{3a}=(C_{ee}^A)^{a3}=(C_{ee}^A)^{3a}=0$ and $(C_{dd}^A)^{33}=(C_{ee}^A)^{33}=1$.
Therefore, all flavor-violating couplings will be suppressed by the light quark and lepton masses.
Now, we will discuss the predictions for the most relevant lepton violating processes and meson decays to show that the experimental bounds can be satisfied.

\begin{itemize}
\item $K^0\!-\!\bar{K}^0$ mixing:

\begin{figure}[t]
\centering
\includegraphics[width=0.475\linewidth]{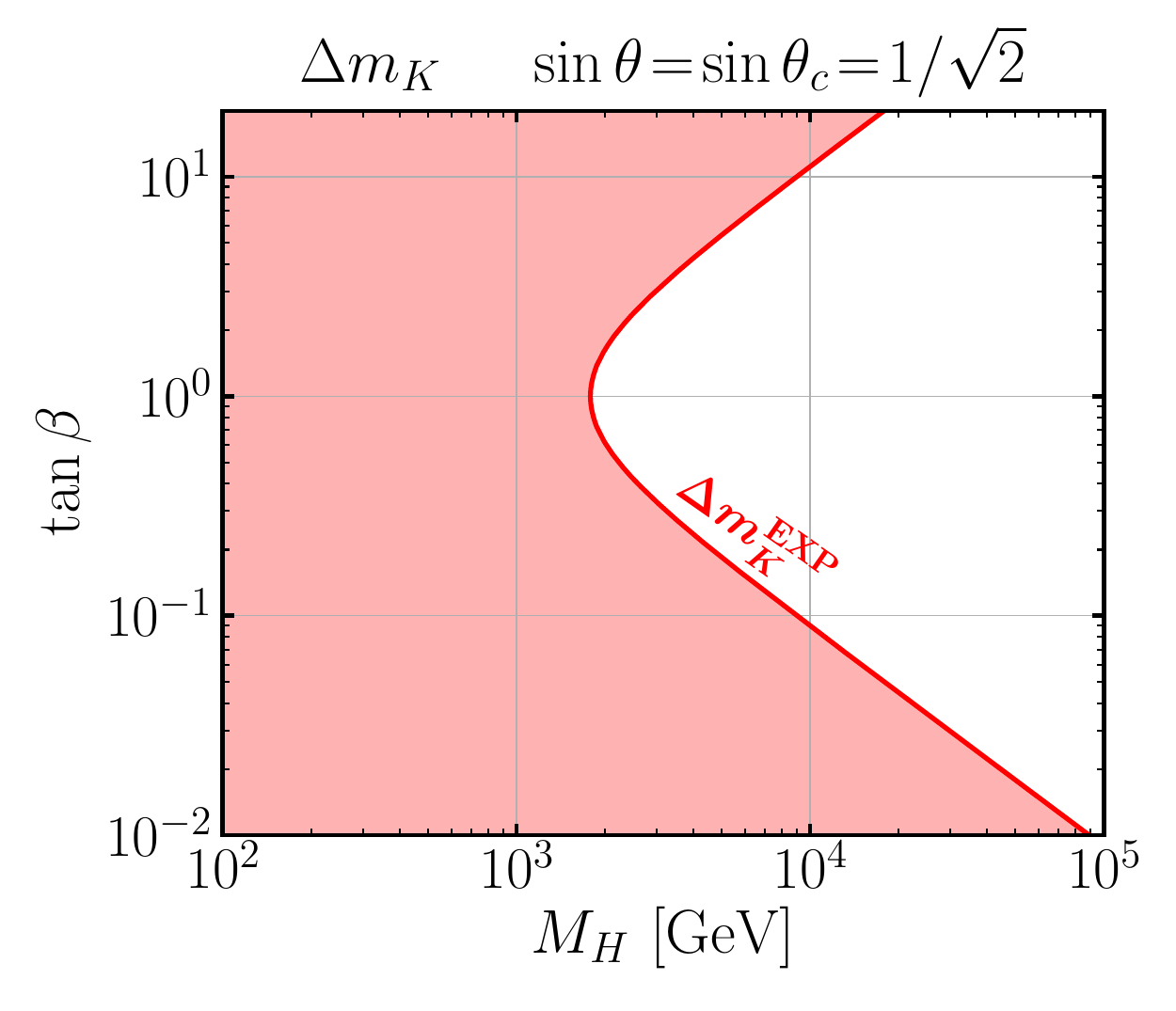}
\includegraphics[width=0.475\linewidth]{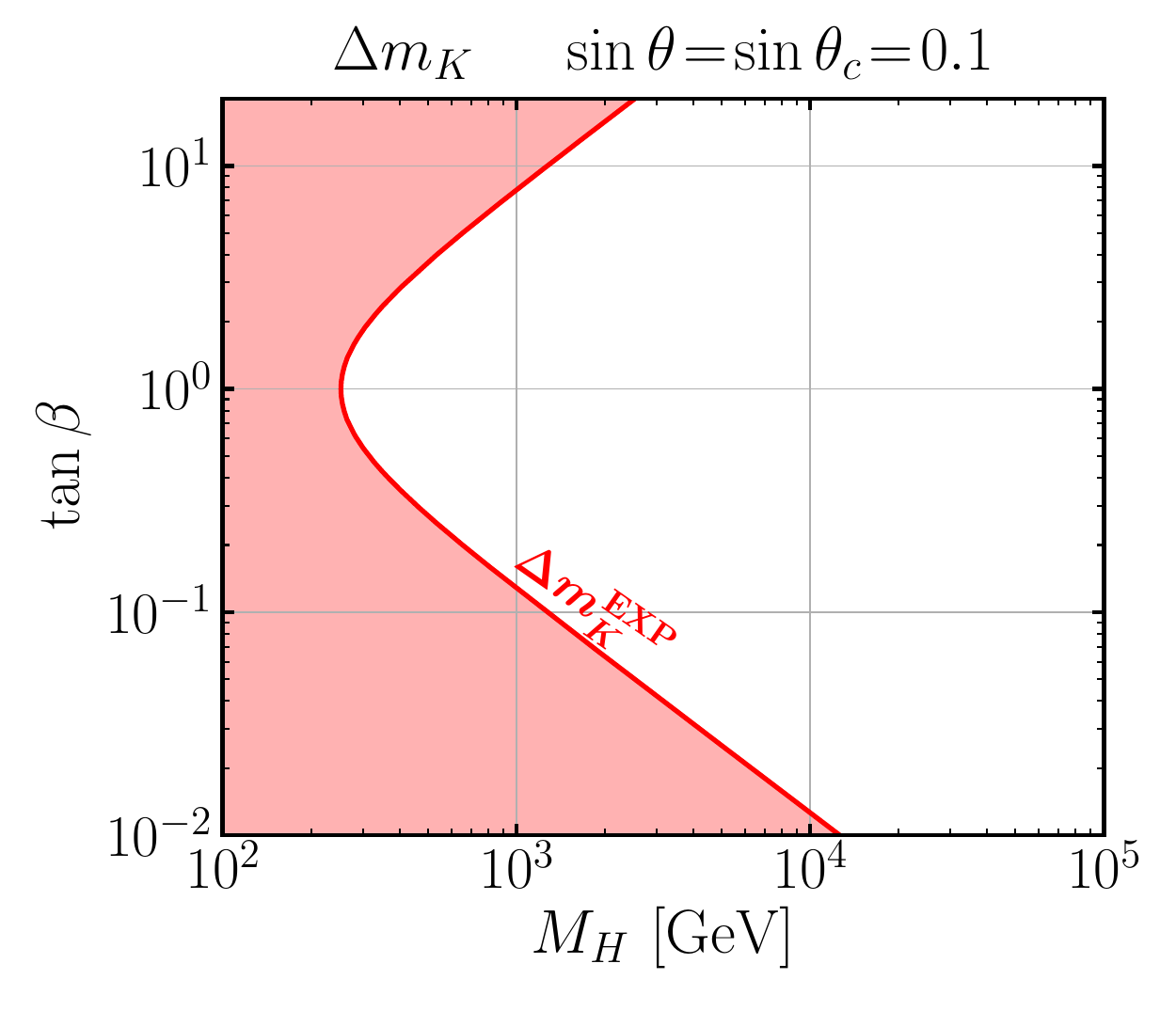}
\caption{Contour plot in the $\tan \beta$ vs $M_H$ plane showing the allowed parameter space by the experimental measurement of $K^0-\bar{K}^0$ mixing. The region shaded in red gives a larger contribution than the measured value of $\Delta m_K^{\rm EXP}$. For the left (right) panel we fix the mixing angles to $\sin \theta= \sin \theta_c=1/\sqrt{2}$ ($\sin \theta= \sin \theta_c=0.1$).}
\label{fig:kcontour}
\end{figure}

\begin{figure}[b]
\centering
\includegraphics[scale=0.75]{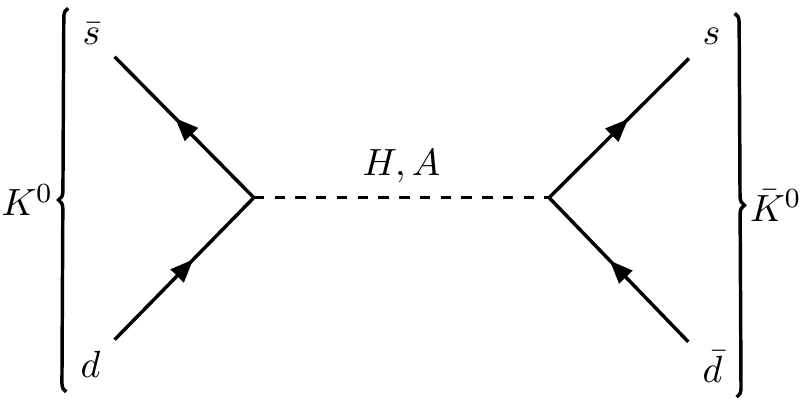}
\caption{Feynman diagram for the contribution from the new scalars $H$ and  $A$ to $K^0-\bar{K}^0$ mixing.}
\label{fig:diagkmixing}
\end{figure}

The flavor-violating couplings in the quark sector will contribute to the measured mass splitting for the $K$ mesons (see diagram in Fig.~\ref{fig:diagkmixing}). For studies of the general 2HDM and meson mixing see e.g. Refs.~\cite{Atwood:1996vj,Babu:2018uik}. We require this contribution to be smaller than the experimental measured value of $\Delta m_K^{\rm EXP}= (3.484\pm 0.006) \times 10^{-15}$ GeV ~\cite{ParticleDataGroup:2018ovx}. For the transition matrix element we use the results presented in Ref.~\cite{Deshpande:1993py}
\begin{align}
M_{12}^K =  - \frac{f_K^2 m_K}{2 M_H^2} & \left[ -\frac{5}{24} \frac{m_K^2}{(m_{q_i} + m_{q_j})^2} \left( {C_{ij}^H}^2 + {C_{ji}^{H*}}^2 \right) B_2 \eta_2(\mu) \right. \nonumber \\
 & \left. +  \, C_{ij}^H C_{ji}^{H*}  \left( \frac{1}{12} + \frac{1}{2} \frac{m_K^2}{(m_{q_i} + m_{q_j})^2}  \right) B_4 \eta_4(\mu)    \right],
\end{align}
where we use $B_2(\mu)=0.66$, $B_4(\mu)=1.03$, $\eta_2(\mu)=2.552$ and $\eta_4(\mu)=4.362$ at $\mu=2$ GeV~\cite{Ciuchini:1998ix,Babu:2018uik}. The $C^H_{ij}$ coefficients are given in Appendix~\ref{sec:appFR}. The mass difference corresponds to $\Delta m_K = 2{\rm Re}\left[M_{12}^K\right]$. The pseudoscalar $A$ gives the same contribution simply by replacing $H \leftrightarrow A$ above.

In Fig.~\ref{fig:kmixing} we present our results for the mass splitting $\Delta m_K$ in units of GeV as a function of the mass of the Higgs bosons where we set $M_H\!=\!M_A$. We take $m_K=498$ MeV and $f_K=160$ MeV for the decay constant. In the left panel we take maximal mixing $\sin \theta\!=\!\sin \theta_c\!=\!1/\sqrt{2}$. The blue line corresponds to $\tan \beta\!=\!1/20$ and the experimental measurement of $\Delta m_K$ requires $M_H \gtrsim 18$ TeV; while for $\tan \beta\!=\!1$ ($\tan \beta\!=\!10$) we find that $M_H \gtrsim 1.8$ TeV ($M_H \gtrsim 9$ TeV) is allowed. This bound becomes weaker as the mixing angles are reduced, this can be seen in the left panel in Fig.~\ref{fig:kmixing}, where we present our results for $\sin \theta\!=\!\sin \theta_c\!=\!0.1$.

In Fig.~\ref{fig:kcontour} we present our results as a contour plot in the $\tan\beta$ vs $M_H$ plane. The region shaded in red is ruled out since it gives a larger contribution than the measured value of $\Delta m_K^{\rm EXP}$. In the left panel we take maximal mixing angles which requires $M_H\gtrsim 1.8$ TeV but for large and small values of $\tan\beta$ to be allowed it requires $M_H \gtrsim 10$ TeV. Therefore, for the Higgs bosons to be around the TeV scale this bound requires the Yukawa couplings to be very close to flavor-diagonal. On the right panel we take small mixing angles of $\sin \theta=\sin \theta_c=0.1$ which require $M_H \gtrsim 250$ GeV.

The couplings will also induce the decay $K_L \to \mu^\pm e^\mp$ which for couplings of $\mathcal{O}(1)$ excludes masses of $10^3$ TeV~\cite{Valencia:1994cj}. However, in this case the four-fermion effective interaction is suppressed by $Y_{ds}Y_{\mu e} \approx m_\mu m_s /(16 v^2) \approx 10^{-8}$, and hence, this bound is much weaker than the one from $\Delta m_K$.

\begin{figure}[t]
\centering
\includegraphics[width=0.475\linewidth]{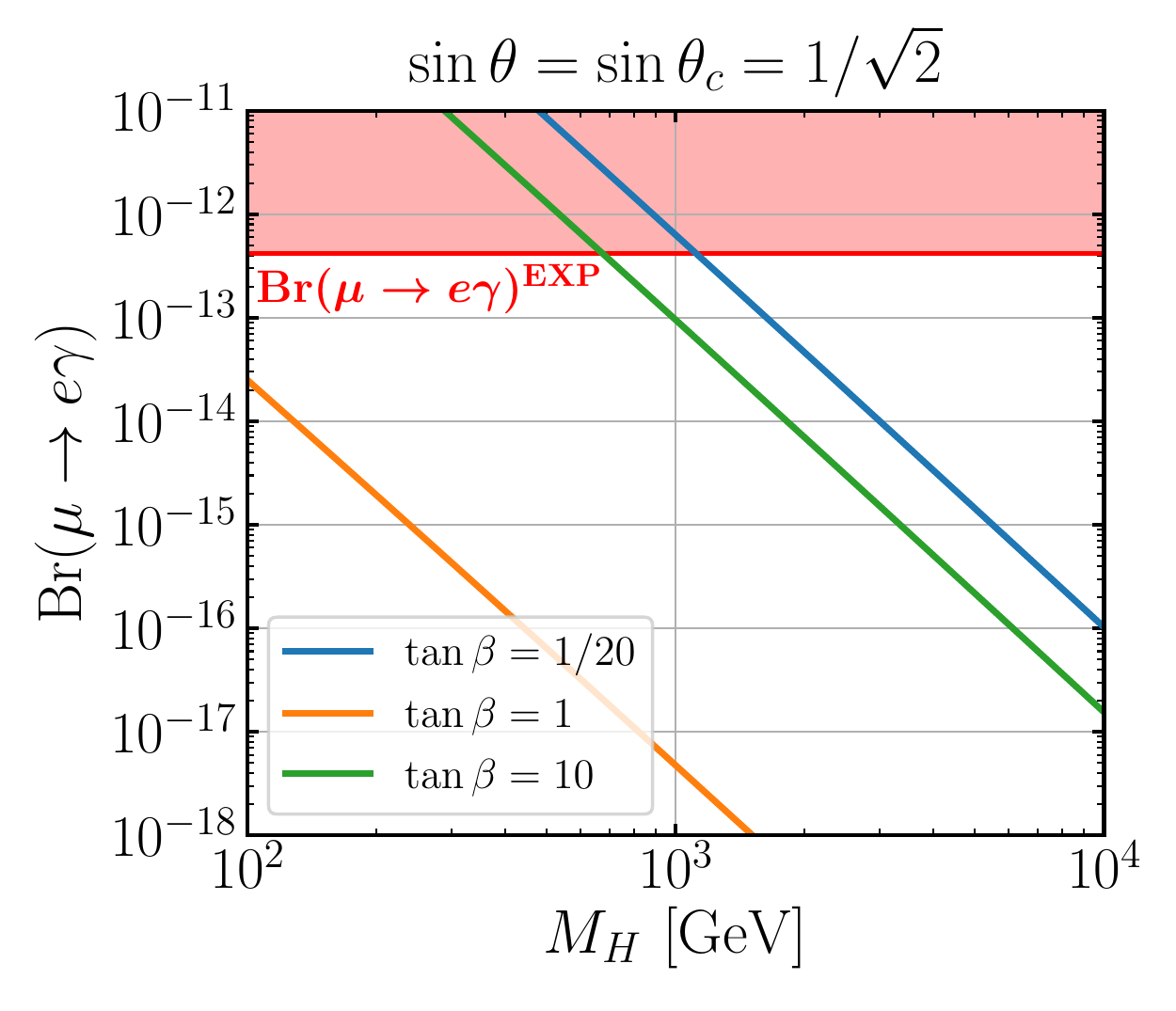}
\includegraphics[width=0.475\linewidth]{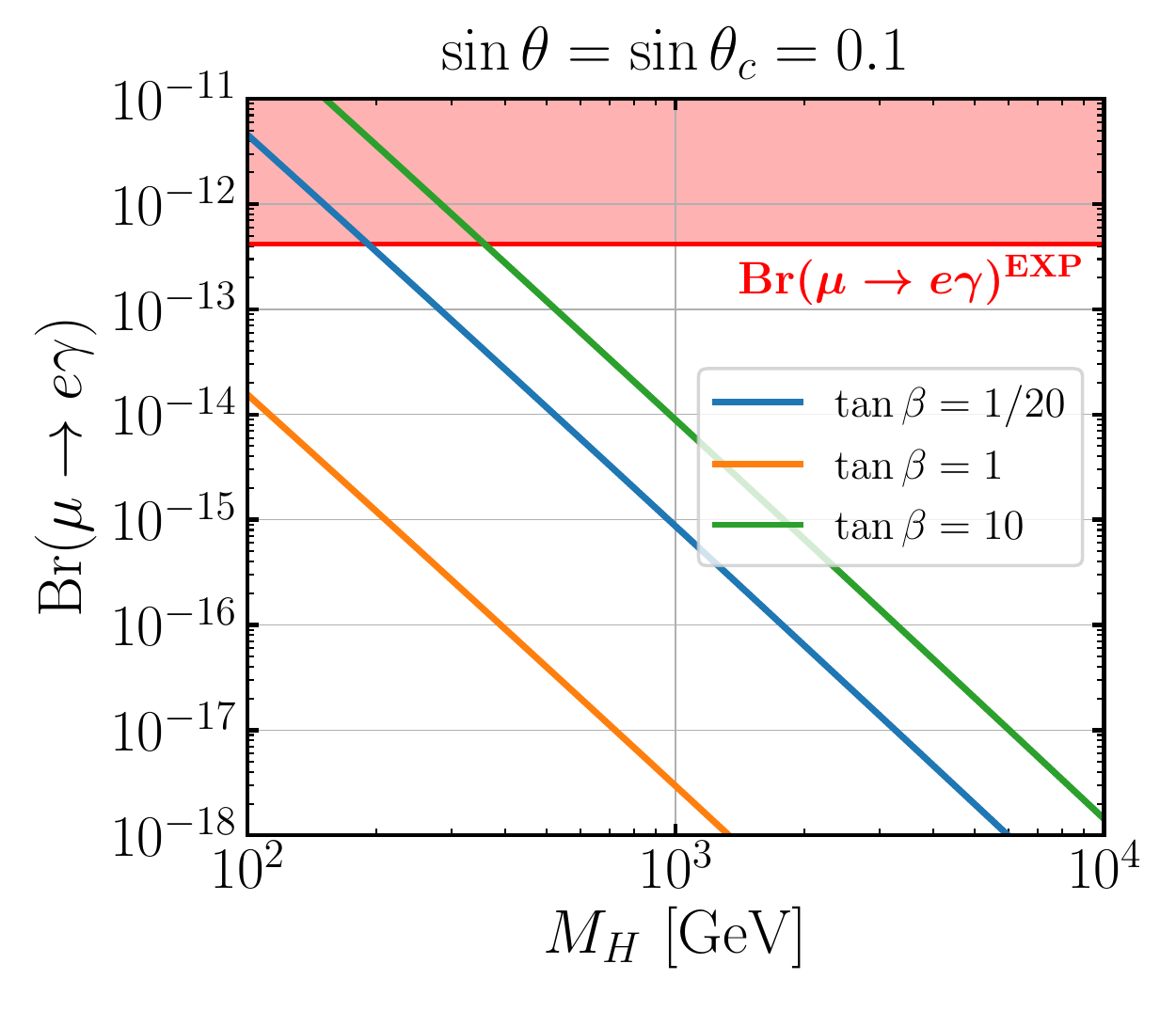}
\caption{Branching ratio for the process $\mu\to e \gamma$ as a function of the mass of the Higgs bosons $M_H=M_A$. The region shaded in red corresponds to the experimental upper bound on this branching ratio. For the left (right) panel we fix the mixing angles to $\sin \theta= \sin \theta_c=1/\sqrt{2}$ ($\sin \theta= \sin \theta_c=0.1$).}
\label{fig:BRmeg}
\end{figure}

\begin{figure}[b]
\centering
\includegraphics[scale=0.75]{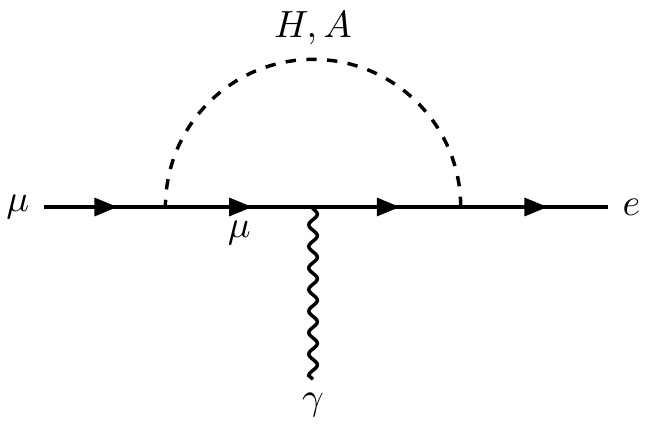}
\caption{Feynman diagram for the contribution from the new scalars $H$ and $A$ to $\mu \to e \gamma$.}
\label{fig:meg}
\end{figure}

\item $\mu \to e \gamma$:

In the leptonic sector, the new neutral scalars can give rise to $\mu \to e \gamma$ at one-loop (see diagram in Fig.~\ref{fig:meg}). Since the coupling to the top-quark is not predicted by the theory and the $\kappa$ parameter can be small, the two-loop Barr-Zee contribution (with the top quark running in the loop) is subleading. This process has been constrained experimentally by the MEG collaboration to be ${\rm Br}(\mu \to e \gamma)\leq 4.2 \times 10^{-13}$~\cite{MEG:2016leq}. For the calculation of this branching ratio we adapt the results in Ref.~\cite{Chang:1993kw},
\beq
{\rm Br} (\mu \to e \gamma) = \frac{3  \alpha }{16 \pi G_F^2} \left( |A_L|^2 + |A_R|^2 \right), 
\eeq
where
\begin{align}
A_L & =  \sum_{S=H,A} \frac{C^S_{\mu\mu} C^S_{e\mu} }{M_S^2} \left( \ln\frac{m_\mu^2}{M_S^2} + \frac{3}{2} \right) , \\[1ex]
A_R & =  \sum_{S=H,A} \frac{C^{S*}_{\mu\mu} C^{S*}_{\mu e} }{M_S^2} \left( \ln\frac{m_\mu^2}{M_S^2} + \frac{3}{2} \right). 
\end{align}
In Fig.~\ref{fig:BRmeg} we present our results for ${\rm Br}(\mu \to e \gamma) $ as a function of the mass of the Higgs bosons and we take $M_H=M_A$. The region shaded in red corresponds to the upper bound reported by the MEG collaboration. The figure on the left corresponds to maximal mixing $\sin\theta=\sin\theta_C=1/\sqrt{2}$. The lines of different colors correspond to different values for $\tan\beta$. The blue (green) line corresponds to $\tan \beta=1/20$ for which this bound requires $M_H\gtrsim 1.1$ TeV ($M_H\gtrsim 700$ GeV ). This experimental bound becomes weaker as we decrease the value of the mixing angle, for the panel on the right we show the same result but for $\sin\theta=\sin\theta_C=0.1$.

In the context of the general 2HDM, the experimental bounds coming from $\mu \to e e e$ and $\mu - e$ conversion have been shown to be subleading to the bound from $\mu \to e \gamma$ ~\cite{Diaz:2000cm, Paradisi:2006jp}.

\end{itemize}
In summary, the bound that comes from the measurement of $\Delta m_K$ is much stronger than the constraint from $\mu \to e\gamma$ on the mixing angles $\theta$ and $\theta_c$. For the Higgs bosons to be around the electroweak scale this requires the Yukawa interactions to be very close to flavor-diagonal which justifies the approach taken in Section~\ref{sec:decays}.
\section{SUMMARY}
\label{sec:summary}
%
Quark-lepton unification remains one of the best-motivated ideas for physics beyond the Standard Model. In this article, we studied the phenomenology of the 2HDM in the minimal theory of quark-lepton unification that can live at the low scale.  In the limit with no flavor-violating couplings we gave concrete predictions for the branching ratios of the heavy neutral and charged Higgs bosons. Moreover, 
we derived relations between the decay widths of the heavy Higgs bosons into quarks and leptons that arise from quark-lepton unification. Namely, for small $\tan \beta$ the decay width into bottom quarks should be three times the decay width into tau leptons, while for large $\tan \beta$ it is the opposite.

We also studied the production cross-sections of the new scalars at the LHC. The current experimental bounds by the ATLAS collaboration already exclude some regions in the parameter space and the future high-luminosity stage will be able to probe this scenario in the TeV regime. In this theory, the cross-section for the processes $pp \to H,A \to \bar{\tau}\tau$ is related to $pp \to H,A \to \bar{b} b$ for small and large values of $\tan\beta$ and can be used in the future to probe the idea of quark-lepton unification.

Furthermore, we studied the experimental constraints on the flavor-violating couplings in the quark and leptonic sectors induced by the Higgs bosons. We demonstrated that the experimental measurement of meson mixing gives strong constraints on the off-diagonal entries and for Higgs bosons around the TeV scale this requires the interacting matrices to be very close to flavor-diagonal. If new scalars beyond the Standard Model Higgs are discovered in the near future, this study provides a path to infer whether the underlying theory arises from quark-lepton unification.\\ 

{\small{\textit{Acknowledgments:}
A.D.P. is  supported by the INFN ``Iniziativa Specifica''   Theoretical Astroparticle Physics  (TAsP-LNF) and by the Frascati National Laboratories (LNF) through a Cabibbo Fellowship,
call 2020. A.D.P. would also like to thank KITP at UC Santa Barbara for their hospitality while this work was being completed with support from the National Science Foundation under Grant No. NSF PHY-1748958.}}

\appendix

\section{Feynman Rules}
\label{sec:appFR}
%
The Feynman for the physical scalars in the two Higgs doublets:
\begin{align}
&\bar{u}^i d^j H^+: \hspace{0.5cm} i \left[ (C_{Lud})^{ij} P_L + (C_{Rud})^{ij} P_R \right], \hspace{1cm}
\bar{d}^i u^j H^-: \hspace{0.5cm} i \left[ (C_{Rud}^*)^{ji} P_L + (C_{Lud}^*)^{ji} P_R \right],  \nonumber \\[1.5ex]
& \bar{\nu}^i e^j H^+ : \hspace{0.5cm} - i ( C_{\nu e} )^{ij} P_R, \hspace{3.8cm}
\bar{N}^i e^j H^+ : \hspace{0.5cm}  i ( C_{Ne} )^{ij} P_L , \nonumber \\[1.5ex]
&\bar{e}^i N^j H^-: \hspace{0.5cm}  i (C_{Ne}^*)^{ji} P_R, \hspace{3.85cm}
\bar{e}^i \nu^j H^-: \hspace{0.5cm}  -i (C_{\nu e}^*)^{ji} P_L, \nonumber \\[1.5ex]
& \bar{u}^i u^j h: \hspace{0.5cm}  i \left[ (C_{uu}^h)^{ij} P_L + (C_{uu}^{h*})^{ji} P_R \right], \hspace{1.7cm}
\bar{N}^i \nu^j h: \hspace{0.5cm}  i \left[ (C_{N\nu}^h)^{ij} P_L + (C_{N\nu}^{h*})^{ji} P_R \right], \nonumber \\[1.5ex]
& \bar{d}^i d^j h: \hspace{0.5cm}  i \left[ (C_{dd}^h)^{ij} P_L + (C_{dd}^{h*})^{ji} P_R \right], \hspace{1.8cm}
\bar{e}^i e^j h: \hspace{0.5cm}  i \left[ (C_{ee}^h)^{ij} P_L + (C_{ee}^{h*})^{ji} P_R \right], \nonumber \\[1.5ex]
&\bar{u}^i u^j H: \hspace{0.5cm}  i \left[ (C_{uu}^H)^{ij} P_L + (C_{uu}^{H*})^{ji} P_R \right],\hspace{1.6cm} 
\bar{N}^i \nu^j H: \hspace{0.5cm}  i \left[ (C_{N\nu}^H)^{ij} P_L + (C_{N\nu}^{H*})^{ji} P_R\right], \nonumber \\[1.5ex]
&\bar{d}^i d^j H: \hspace{0.5cm}  i \left[ (C_{dd}^H)^{ij} P_L + (C_{dd}^{H*})^{ji} P_R \right],\hspace{1.7cm}
\bar{e}^i e^j H: \hspace{0.5cm}  i \left[ (C_{ee}^H)^{ij} P_L + (C_{ee}^{H*})^{ji} P_R \right], \nonumber \\[1.5ex]
&\bar{u}^i u^j A: \hspace{0.5cm}   (C_{uu}^A)^{ij} P_L - (C_{uu}^{A*})^{ji} P_R\,, \hspace{2.2cm}
\bar{N}^i \nu^j A: \hspace{0.5cm}   (C_{N\nu}^A)^{ij} P_L - (C_{N\nu}^{A*})^{ji} P_R\, , \nonumber \\[1.5ex]
&\bar{d}^i d^j A: \hspace{0.5cm}   (C_{dd}^A)^{ij} P_L - (C_{dd}^{A*})^{ji} P_R\, , \hspace{2.25cm}
\bar{e}^i e^j A: \hspace{0.5cm}   (C_{ee}^A)^{ij} P_L - (C_{ee}^{A*})^{ji} P_R\, , \nonumber \\[1.5ex]
& Z Z h: \hspace{0.5cm} i \frac{v \, g_2^2 \sin(\beta -\alpha)}{8\cos^2{\theta_W}},\hspace{3.6cm}
W^+ W^- h: \hspace{0.5cm} i \frac{v \, g_2^2 \sin(\beta -\alpha)}{4\cos^2{\theta_W}}, \nonumber \\[1.5ex]
&Z Z H: \hspace{0.5cm} i \frac{v \, g_2^2 \cos(\beta -\alpha)}{8\cos^2{\theta_W}},\hspace{3.4cm}
W^+ W^- H: \hspace{0.5cm} i \frac{v \, g_2^2 \cos(\beta -\alpha)}{4\cos^2{\theta_W}}, \nonumber \\[1.5ex]
& h(p_1) A(p_2) Z:  \hspace{0.5cm}  i \frac{g_2 \cos(\beta-\alpha)}{2\cos \theta_W} (p_1-p_2)_\mu, \nonumber \\[1.5ex]
& H(p_1) A(p_2) Z:  \hspace{0.5cm}  i \frac{g_2 \sin(\beta-\alpha)}{2\cos \theta_W} (p_2-p_1)_\mu, \nonumber \\[1.5ex]
& H^\pm(p_1) h(p_2) W^\mp:   \hspace{0.5cm} i \frac{g_2}{2} \cos(\beta-\alpha) (p_2-p_1)_\mu, \nonumber \\[1.5ex]
& H^\pm(p_1) H(p_2) W^\mp:  \hspace{0.5cm}  i \frac{g_2}{2} \sin(\beta-\alpha) (p_1-p_2)_\mu, \nonumber \\[1.5ex]
&HHH: \hspace{0.5cm} \frac{v}{2} \left(\lambda_1 c^3_{\alpha} c_{\beta} + \lambda_2 s^3_{\alpha} s_{\beta} + (\lambda_{3} + \lambda_{4}) s_{\alpha}c_{\alpha} s_{\alpha + \beta}\right), \nonumber \\[1.5ex]
&hhh: \hspace{0.5cm}  \frac{v}{2} \left(-\lambda_1 s^3_{\alpha} c_{\beta} + \lambda_2 c^3_{\alpha} s_{\beta} - (\lambda_{3} + \lambda_{4}) s_{\alpha}c_{\alpha} c_{\alpha + \beta}\right), \nonumber \\[1.5ex]
&Hhh: \hspace{0.5cm}
v \left( 3 \lambda_1 c_\beta c_\alpha s_\alpha^2 +3 \lambda_2 s_\beta c_\alpha^2 s_\alpha + \lambda_{345} (c_\beta c_\alpha^3 - 2 c_\beta c_\alpha s_\alpha^2 - 2 s_\beta c^2_\alpha s_\alpha + s_\beta s_\alpha^3)  \right. \nonumber \\[1.5ex]
&\hspace{2cm} \left. 3 \lambda_6 (c_\beta s^3_\alpha + s_\beta c_\alpha s_\alpha^2 - 2 c_\beta c^2_\alpha s_\alpha ) +3\lambda_7(s_\beta c^3_\alpha + c_\beta c^2_\alpha s_\alpha - 2 s_\beta c_\alpha s^2_\alpha)\right), \nonumber \\[1.5ex]
&hHH: \hspace{0.5cm}
\frac{v}{2} ( -3 \lambda_1 c_\beta c_\alpha^2 s_\alpha + 3 \lambda_2 s_\beta c_\alpha s_\alpha^2 + \lambda_{345}(s_\beta c_\alpha^3 + 2c_\beta c_\alpha^2 s_\alpha - 2 s_\beta c_\alpha s_\alpha^2 - c_\beta s_\alpha^3) \nonumber \\[1.5ex]
&\hspace{2cm} + 3 \lambda_6 (c_\beta c_\alpha^3 - s_\beta c_\alpha^2 s_\alpha - 2 c_\beta c_\alpha s_\alpha^2) + 3 \lambda_7(2s_\beta c_\alpha^2 s_\alpha + c_\beta c_\alpha s_\alpha^2 - s_\alpha^3)), \nonumber
\end{align}
\begin{align}
&HAA: \hspace{0.5cm}
v \left( -\lambda_1 c_\beta s_\beta^2 c_\alpha + \lambda_2 c_\beta^2 s_\beta s_\alpha + (\lambda_{3} + \lambda_4) (c_\beta^3 c_\alpha  + s_\beta^3 s_\alpha ) + \frac{\lambda_5}{4} (c_{\alpha + 3\beta} - 5c_{\alpha - \beta}) \right. \nonumber \\[1.5ex]
&\hspace{2cm} \left. \frac{\lambda_6}{2} s_\beta (c_\alpha + c_{\alpha-2\beta} + 2 c_{\alpha + 2 \beta} ) + \frac{\lambda_7}{2} c_\beta ( - s_\alpha +s_{\alpha - 2\beta}+ 2 s_{\alpha + 2 \beta})\right), \nonumber \\[1.5ex]
&hAA: \hspace{0.5cm}
v \left( - \lambda_1 c_\beta s_\beta^2 s_\alpha + \lambda_2 c_\beta^2 s_\beta c_\alpha + (\lambda_{3} + \lambda_4) ( - c_\beta^3 s_\alpha  + s_\beta^3 c_\alpha ) + \frac{\lambda_5}{4} (s_{\alpha + 3\beta} - 5s_{\alpha - \beta}) \right. \nonumber \\
&\hspace{2cm} \left. \frac{\lambda_6}{2} s_\beta (s_\alpha + s_{\alpha-2\beta} + 2 s_{\alpha + 2 \beta} ) + \frac{\lambda_7}{2} c_\beta ( - c_\alpha + c_{\alpha - 2\beta}+ 2 c_{\alpha + 2 \beta})\right), \nonumber \\
&HH^+H^-: \hspace{0.5cm}
v \left( \lambda_1 c_\beta s_\beta^2 c_\alpha + \lambda_2 c_\beta s_\beta^2 s_\alpha + \lambda_{3} s_\beta (c_\beta^2 c_\alpha  + s_\beta^2 s_\alpha ) + \frac{(\lambda_4 + \lambda_5)}{4} s_{\alpha + \beta}(1 - c_{2\beta} + s_{2\beta}) \right. \nonumber \\
&\hspace{2cm} +\left. \frac{\lambda_6}{2} s_\beta (3c_\alpha + c_{\alpha-2\beta}) + \lambda_7 s_\beta ( s_\alpha + c_\beta s_\beta (c_\alpha + s_\alpha))\right), \nonumber \\
&hH^+H^-: \hspace{0.5cm}
v \left( -\lambda_1 c_\beta s_\beta^2 s_\alpha + \lambda_2 c_\beta s_\beta^2 c_\alpha + \lambda_{3} s_\beta (-c_\beta^2 c_\alpha  + s_\beta^2 s_\alpha ) + \frac{(\lambda_4 + \lambda_5)}{4} c_{\alpha + \beta}(1 - c_{2\beta} + s_{2\beta}) \right. \nonumber \\
&\hspace{2cm} - \left. \frac{\lambda_6}{2} s_\beta (3s_\alpha + s_{\alpha-2\beta}) + \lambda_7 s_\beta ( c_\alpha + c_\beta s_\beta (c_\alpha - s_\alpha))\right), \nonumber 
\end{align}
where the interaction matrices are given by
\begin{align}
C_{Lud} & = U_c^T \left( Y_1^T \sin \beta  - Y_2^T \frac{\cos \beta}{2 \sqrt{3}} \right) D, \hspace{1.4cm}
C_{Rud} = - U^\dagger \left( Y_3^* \sin \beta -Y_4^*  \frac{\cos \beta}{2 \sqrt{3}} \right) D_c^* , \nonumber \\[1.5ex]
C_{Ne} & = N_c^T \left( Y_1^T \sin \beta + Y_2^T \frac{\sqrt{3} \cos \beta}{2} \right) E, \hspace{0.8cm}
C_{\nu e} = N^\dagger \left(  Y_3^* \sin \beta  + Y_4^* \frac{\sqrt{3} \cos \beta}{2} \right) E_c^*, \nonumber \\[1.5ex]
C_{uu}^H & = U^T_c \left( Y_1^T \frac{\cos \alpha}{\sqrt{2}} + Y_2^T \frac{\sin \alpha}{ 2\sqrt{6}} \right)   U , \hspace{1.4cm}
C_{N\nu}^H = N_c^T \left(  Y_1^T \frac{\cos \alpha}{\sqrt{2}} - Y_2^T \frac{3 \sin \alpha}{2\sqrt{6}} \right) N , \nonumber \\[1.5ex]
C_{dd}^H & = D_c^T \left( Y_3^T \frac{\cos \alpha}{\sqrt{2}} + Y_4^T \frac{\sin \alpha}{ 2 \sqrt{6}}  \right) D , \hspace{1.3cm}
C_{ee}^H = E_c^T \left( Y_3^T \frac{\cos \alpha}{\sqrt{2}} - Y_4^T \frac{3\sin \alpha}{2\sqrt{6}}  \right) E, \nonumber \\[1.5ex]
C_{uu}^h & = U^T_c \left( -Y_1^T \frac{\sin \alpha}{\sqrt{2}} + Y_2^T \frac{\cos \alpha}{ 2 \sqrt{6}} \right)   U , \hspace{1.1cm}
C_{N\nu}^h = N_c^T \left( -Y_1^T \frac{\sin \alpha}{\sqrt{2}} - Y_2^T \frac{3 \cos \alpha}{2\sqrt{6}} \right) N , \nonumber \\[1.5ex]
C_{dd}^h & = D_c^T \left( -Y_3^T \frac{\sin \alpha}{\sqrt{2}} + Y_4^T \frac{\cos \alpha}{ 2\sqrt{6}}  \right) D , \hspace{1cm}
C_{ee}^h = E_c^T \left( -Y_3^T \frac{\sin \alpha}{\sqrt{2}} - Y_4^T \frac{3\cos \alpha}{2\sqrt{6}}  \right) E, \nonumber \\[1.5ex]
C_{uu}^A & = U_c^T \left(Y_1^T \frac{\sin \beta}{\sqrt{2}} - Y_2^T \frac{\cos \beta}{2 \sqrt{6}}  \right) U ,\hspace{1.43cm}
C_{N\nu}^A = N_c^T \left(Y_1^T \frac{\sin\beta}{\sqrt{2}} + Y_2^T \frac{3\cos\beta}{2\sqrt{6}} \right)  N , \nonumber \\[1.5ex]
C_{dd}^A & = D_c^T \left( - Y_3^T \frac{\sin \beta}{\sqrt{2}} + Y_4^T \frac{\cos \beta}{  2 \sqrt{6}}  \right) D  , \hspace{1.1cm}
C_{ee}^A = E_c^T \left( - Y_3^T \frac{\sin \beta}{\sqrt{2}} - Y_4^T \frac{3 \cos \beta}{ 2 \sqrt{6}}  \right) E . \nonumber
\end{align}
In the alignment limit $\sin(\beta-\alpha)\simeq 1$, the Higgs interactions take the following form
\begin{align}
&HHH: \hspace{0.5cm} - i g_2 \frac{3}{2}\frac{M_H^2}{M_W}, \hspace{10cm}\nonumber
\end{align}
\begin{align}
&hhh: \hspace{0.9cm} - i g_2 \frac{3}{2}\frac{M_h^2}{M_W}, \nonumber \\[1ex]
&Hhh: \hspace{0.9cm} i \frac{3v}{2} \left( \lambda_1s_{2\beta}(c_{2\beta}+1) + \lambda_2s_{2\beta}(c_{2\beta}-1) - 2 \lambda_{345}c_{2\beta}s_{2\beta} - 2 \lambda_6(c_{2\beta} + c_{4\beta}) \right. \nonumber\\ &\hspace{2.5cm} \left. - 2\lambda_7(c_{2\beta} - c_{4\beta}) \right) = i \lambda_{\rm eff}, \nonumber \\[1ex]
&hHH: \hspace{0.9cm} i\frac{v}{8} \left( 3\lambda_1 (1 - c_{4\beta}) + 3 \lambda_2 (1 - c_{4\beta}) + 2 \lambda_{345} (1+3c_{4\beta}) + 12(\lambda_7 - \lambda_6) s_{4\beta} \right), \nonumber \\[1ex]
&hH^+H^-: \hspace{0.5cm} i\frac{v}{8} \left( 8 \lambda_1 c^2_{\beta} s^2_{\beta} + 8 \lambda_2 c_{\beta} s^3_{\beta} + \lambda_3 (3 - 4 c_{2\beta} + c_{4\beta} + 2 s_{2\beta} + s_{4\beta}) \right. \nonumber \\ 
&\hspace{2.5cm} \left. + 4(\lambda_4 + \lambda_5) s_{\beta}^2(1 + c_{2\beta} + s_{2 \beta})  + \lambda_6 s_{2\beta} + 4\lambda_7 s^2_{\beta} (3 + c_{2\beta} + s_{2\beta}) \right) . \nonumber
\end{align}
The interaction matrices in terms of the physical fermion masses are given by
\begin{align}
C_{dd}^H & = \left( \frac{\sin{\alpha}}{\sin{\beta}} + 3\frac{\cos{\alpha}}{\cos{\beta}}  \right) \frac{M_D^{\rm diag}}{4v} + \left( \frac{\cos{\alpha}}{\cos{\beta}} - \frac{\sin{\alpha}}{\sin{\beta}}\right)\frac{V_c^{*} M_E^{\rm diag} V^{\dagger}}{4 v}, \nonumber \\[1.5ex]
C_{ee}^H & = \left( \frac{\cos{\alpha}}{\cos{\beta}} + 3\frac{\sin{\alpha}}{\sin{\beta}}  \right) \frac{M_E^{\rm diag}}{4v} + 3\left( \frac{\cos{\alpha}}{\cos{\beta}} -\frac{\sin{\alpha}}{\sin{\beta}}\right)\frac{V_c^{T} M_D^{\rm diag} V}{4 v}, \nonumber \\[1.5ex]
C_{uu}^H & = \left( 3\frac{\cos{\alpha}}{\cos{\beta}} + \frac{\sin{\alpha}}{\sin{\beta}}  \right) \frac{M_U^{\rm diag}}{4v} + \left( \frac{\cos{\alpha}}{\cos{\beta}} - \frac{\sin{\alpha}}{\sin{\beta}}\right)\frac{ U_c^T M_\nu^{D \,T} U }{4 v}, \nonumber \\[1.5ex]
C_{N\nu}^H & = \left( \frac{\cos{\alpha}}{\cos{\beta}} + 3\frac{\sin{\alpha}}{\sin{\beta}}  \right) \frac{N_c^T{M_\nu^{D \,T}} N}{4v} + 3\left( \frac{\cos{\alpha}}{\cos{\beta}} - \frac{\sin{\alpha}}{\sin{\beta}}\right)\frac{V_1^{*} M_U^{\rm diag} K_1 V_{\rm CKM} K_2 V K_3 V_{\rm PMNS}}{4 v}, \nonumber \\[1.5ex]
C_{dd}^A & = \left( \cot{\beta} - 3\tan{\beta}  \right) \frac{M_D^{\rm diag}}{4v} - \left( \tan{\beta} + \cot{\beta}\right)\frac{V_c^{*} M_E^{\rm diag} V^{\dagger}}{4 v}, \nonumber \\[1.5ex]
C_{ee}^A & = \left( 3\cot{\beta} - \tan{\beta}  \right) \frac{M_E^{\rm diag}}{4v} - 3\left( \tan{\beta} + \cot{\beta}\right)\frac{V_c^{T} M_D^{\rm diag} V}{4 v}, \nonumber \\[1.5ex]
C_{uu}^A & = \left( 3\tan{\beta} - \cot{\beta}  \right) \frac{M_U^{\rm diag}}{4v} + \left( \tan{\beta} + \cot{\beta}\right)\frac{U_c^T M_\nu^{D \,T} U}{4 v},\nonumber \\[1.5ex]
C_{N\nu}^A & = \left(\tan{\beta} - 3\cot{\beta} \right) \frac{N_c^T M_\nu^{D \,T} N}{4 v} + 3 \left( \tan{\beta} + \cot{\beta}\right) \frac{V_1^{*} M_U^{\rm diag} K_1 V_{\rm CKM} K_2 V K_3 V_{\rm PMNS}}{4 v}, \nonumber \\[1.5ex]
C_{Lud} &= (3\tan{\beta} - \cot{\beta})\frac{M_U^{\rm diag} K_1 V_{\rm CKM}K_2}{2 \sqrt{2} v} + (\tan{\beta} + \cot{\beta})\frac{U_c^T {M_{\nu}^{D\,T}} D}{2 \sqrt{2} v}, \nonumber \\[1.5ex]
C_{Rud} &= -(3\tan{\beta} - \cot{\beta})\frac{ K_1 V_{\rm CKM} K_2 M_D^{\rm diag}}{2 \sqrt{2} v} - (\tan{\beta} + \cot{\beta})\frac{ K_1 V_{\rm CKM} K_2 V M_{E}^{\rm diag} V_c^{\dagger}}{2 \sqrt{2} v}, \nonumber \\[1.5ex]
C_{Ne} &= (\tan{\beta} - 3\cot{\beta}) \frac{N_C^T {M_\nu^{D\,T}} E}{2\sqrt{2} v} +3(\tan{\beta} + \cot{\beta}) \frac{V_1^{*} M_U^{\rm diag} K_1 V_{\rm CKM} K_2 V}{2\sqrt{2}v}, \nonumber \\[1.5ex]
C_{\nu e} &= 3(\tan{\beta} + \cot{\beta})\frac{V_{\rm PMNS}^\dagger K_3^{*} V^\dagger M_D^{\rm diag} V_{c}^{*}}{2 \sqrt{2} v} + (\tan{\beta} - 3\cot{\beta})\frac{V_{\rm PMNS}^\dagger K_3^{*} M_E^{\rm diag}}{ 2\sqrt{2} v}. \nonumber
\end{align}
The interaction matrices of $H$ in the Standard Model limit are given by
\begin{align}
C_{dd}^H & = \left( 3\tan{\beta} - \cot{\beta}  \right) \frac{M_D^{\rm diag}}{4v} + \left( \tan{\beta} + \cot{\beta}\right)\frac{V_c^{*} M_E^{\rm diag} V^{\dagger}}{4 v}, \nonumber \\[1.5ex]
C_{ee}^H & = \left( \tan{\beta} - 3\cot{\beta}  \right) \frac{M_E^{\rm diag}}{4v} + 3\left( \tan{\beta} + \cot{\beta}\right)\frac{V_c^{T} M_D^{\rm diag} V}{4 v}, \nonumber \\[1.5ex]
C_{uu}^H & = \left( 3\tan{\beta} - \cot{\beta}  \right) \frac{M_U^{\rm diag}}{4v} + \left( \tan{\beta} + \cot{\beta}\right)\frac{ U_c^T M_\nu^{D \,T} U }{4 v}, \nonumber \\[1.5ex]
C_{N\nu}^H & = \left( \tan{\beta} - 3\cot{\beta}  \right) \frac{N_c^T M_\nu^{D\,T} N }{4v} + 3\left( \tan{\beta} + \cot{\beta}\right)\frac{V_1^{*} M_U^{\rm diag} K_1 V_{\rm CKM} K_2 V K_3 V_{\rm PMNS}}{4 v}. \nonumber
\end{align}
The interaction matrices are given by:
\begin{align}
	V_1 & = N^{\dagger}_c U_c,  \hspace{2.1cm} V_c = D_c^{\dagger} E_c, \hspace{1.7cm} V_3 = U^{T} Y_2 N_c,\\[1ex] 
V_4 & = N^T Y_4 D_c,  \hspace{1.6cm} V_5 = N^{T}Y_2 U_c,\hspace{1.2cm} V_6 = U^{T} Y_4 E_c, \nonumber \\[1ex] 
	U^{\dagger}D & = K_1 V_{\rm CKM} K_2,\hspace{0.7cm} E^{\dagger} N = K_3 V_{\rm PMNS}, \hspace{1.cm} V = D^{\dagger} E, \nonumber
\end{align}
$K_1$ and $K_3$ are diagonal matrices containing three phases, $K_2$ is a diagonal matrix with two phases.

As can be seen from the Feynman rules, due to quark-lepton unification the off-diagonal entries for the quark interactions will depend on the lepton masses and vice versa. For the approximation given above the mixing matrices take the form:
\begin{align}
V_c^{*} M_E^{\rm diag} V^\dagger & = \begin{pmatrix}
m_e \cos \theta \cos \theta_c + m_\mu \sin \theta \sin \theta_c & -m_e \sin \theta \cos \theta_c  + m_\mu  \cos \theta \sin \theta_c & 0\\
-m_e \cos \theta \sin \theta_c + m_\mu \sin\theta \cos \theta_c & m_e \sin \theta \sin \theta_c + m_\mu  \cos\theta \cos \theta_c  & 0 \\
0 & 0 & m_\tau
\end{pmatrix}, \label{eq:ansatz1} \\[2ex]
V^{T}_c M_D^{\rm diag} V & = \begin{pmatrix}
m_d \cos \theta \cos \theta_c + m_s \sin \theta \sin \theta_c & m_d \sin \theta \cos \theta_c - m_s \cos \theta \sin \theta_c & 0\\
m_d \cos \theta \sin \theta_c - m_s \sin \theta \cos \theta_c & m_d \sin \theta \sin \theta_c  + m_s \cos \theta \cos \theta_c & 0 \\
0 & 0 & m_b
\end{pmatrix}, \label{eq:ansatz2}
\end{align}
the above matrices enter in the couplings $C_{dd}$ and $C_{ee}$. Therefore, Eqs.~\eqref{eq:ansatz1} and~\eqref{eq:ansatz2} can be seen as an ansatz for the Yukawa matrices that arises from quark-lepton unification. This is different from the commonly used Cheng-Sher ansatz~\cite{Cheng:1987rs}.

For a general Higgs decay that couples to massive fermions in the following way
\begin{align}
	\phi \overline{f}_j f_i: \hspace{0.5cm} i(A + B \gamma^5), \nonumber
\end{align}
the decay width corresponds to
\begin{align}
\Gamma(\phi \rightarrow \overline {f}_j f_i) = & \frac{N_C}{8 \pi M_{\phi}} \left\{ \abs{A}^2 \left[M_\phi^2 -(M_i+M_j)^2\right] + \abs{B}^2 \left[M_\phi^2 -(M_i-M_j)^2\right] \right\}  \nonumber \\
 & \times \sqrt{1-\frac{(M_i+M_j)^2}{M_\phi^2}} \sqrt{1-\frac{(M_i-M_j)^2}{M_\phi^2}}  ,
\end{align}
where $N_C$ corresponds to the color factor for the fermions.

The numerical values for the SM fermion masses used in our calculations are evaluated at the $\mu\!=\!M_Z$ scale
\begin{align}
	M_t(M_Z) &= 173.2 \pm 2.4 \ {\rm GeV} \nonumber \\
	M_c(M_Z) &= 0.63 \pm 0.08 \ {\rm GeV} \nonumber \\
	M_b(M_Z) &= 2.89 \pm 0.11 \ {\rm GeV} \nonumber \\
	M_s(M_Z) &= 56 \pm 16 \ {\rm MeV} \nonumber \\
	M_{\tau}(M_Z) &= 1746.45^{+0.29}_{-0.26} \ {\rm MeV} \nonumber \\
	M_\mu (M_Z) &= 102.73 \ {\rm MeV}. \nonumber
\end{align}

\section{Masses of scalar fields}
\label{app:higgsmasses}

The minimization conditions (with $\lambda_{345} = \lambda_3 + \lambda_4 + \lambda_5$) read as:
\begin{align}
	&m_{11}^2 v_1 - m_{12}^2 v_2 + \frac{1}{2}\lambda_1 v_1^3 + \frac{1}{2} \lambda_{345} v_1 v_2^2 + \frac{3}{2} \lambda_6 v_1^2 v_2 + \frac{1}{2} \lambda_7 v_2^3= 0, \\
	&m_{22}^2 v_2 - m_{12}^2 v_1 + \frac{1}{2} \lambda_2 v_2^3 + \frac{1}{2} \lambda_{345} v_1^2 v_2 + \frac{1}{2} \lambda_6 v_1^3 + \frac{3}{2} \lambda_7 v_1 v_2^2= 0,
\end{align}
with
\begin{align}
	m_{11}^2 + \frac{3}{2} \lambda_1 v_1^2 + \frac{1}{2} \lambda_{345} v_2^2 + 3 \lambda_6 v_1 v_2 > 0, \\
	m_{22}^2 + \frac{3}{2} \lambda_2 v_2^2 + \frac{1}{2} \lambda_{345} v_1^2 + 3 \lambda_7 v_1 v_2 > 0.
\end{align}


The physical Higgs masses are given by:
\begin{align}
	M_H^2 &= m_{12}^2 \frac{s_{\alpha - \beta}^2}{s_{2\beta}} - \frac{v^2}{2} \left[ 2 \lambda_1 c^2_\alpha c^2_\beta + 2 \lambda_2 s^2_\alpha s^2_\beta + 4 \lambda_{345} c_\alpha c_\beta s_\alpha s_\beta \right. \nonumber\\
	&\hspace{1cm} \left. + \lambda_6 c_\beta(3c_\alpha^2 s_\beta + 3 s_{2\alpha}c_\beta - ct_\beta s_\alpha^2 c_\beta)  + \lambda_7  s_\beta (3 s_\alpha^2 c_\beta + 3 s_{2\alpha} s_\beta - t_\beta c_\alpha^2 s_\beta) \right],\\
	M_h^2 &= m_{12}^2 t_{\beta} (c_{\alpha}ct_{\beta} + s_{\alpha})^2 - \frac{v^2}{2} \left[ 2 \lambda_1 s_\alpha^2 c_\beta^2 + 2 \lambda_2 c_\alpha^2 s_\beta^2 - 4\lambda_{345} c_\alpha c_\beta s_\alpha s_\beta \right. \nonumber\\
	&\hspace{1cm} + \lambda_6 ct_{\beta}(2c_{2(\alpha + \beta)}-c_{2(\alpha - \beta)}-2c_{2\alpha} - 2 c_{2\beta}+1)  \nonumber\\
	&\hspace{1cm} \left. + \lambda_7 t_{\beta}(1 + 2c_{2\beta} + c_{2\alpha}(2 + c_{2\beta}) - 3s_{2\alpha}s_{2\beta} \right], \\
	M_A^2 &= \frac{m_{12}^2}{s_\beta c_\beta} - \frac{v^2}{2} (2\lambda_5 + \lambda_6 ct_{\beta} + \lambda_7 t_{\beta}),\\
	M_{H^{\pm}}^2 &= \frac{m_{12}^2}{s_\beta c_\beta} - \frac{v^2}{2} ( \lambda_4 + \lambda_5 + \lambda_6 ct_{\beta} + \lambda_7 t_{\beta}).
\end{align}
The masses of all the other Higgs fields can now be written in terms of the mass of the CP-odd Higgs:
\begin{align}
M_H^2&= M_A^2 s_{\beta - \alpha}^2 + v^2 \left( \lambda_1 c_\beta^2 c_\alpha^2 + \lambda_2 s_\beta^2 s_\alpha^2 + 2 \lambda_{345} c_\alpha c_\beta s_\alpha s_\beta + \lambda_5 s^2_{\beta - \alpha} \right. \nonumber \\
&\hspace{2.5cm} \left.+ 2 \lambda_6 c_\beta c_\alpha s_{\beta+\alpha} + 2 \lambda_7 s_\beta s_\alpha s_{\beta+\alpha}\right) , \\
M_h^2&= M_A^2 c_{\beta - \alpha}^2 + v^2 \left( \lambda_1 c_\beta^2 s_\alpha^2 + \lambda_2 s_\beta^2 c_\alpha^2 - 2 \lambda_{345} c_\alpha c_\beta s_\alpha s_\beta + \lambda_5 c^2_{\beta - \alpha} \right.  \nonumber \\
&\hspace{2.5cm} \left. - 2 \lambda_6 c_\beta s_\alpha c_{\beta+\alpha} + 2 \lambda_7 s_\beta c_\alpha c_{\beta + \alpha} \right)  ,\\
M_{H^\pm}^2 &= M_A^2 + \frac{v^2}{2} (\lambda_5 - \lambda_4) .
\end{align}

\bibliography{QL-2HDM}

\end{document}